\pdfoutput=1 
\documentclass[12pt]{article}
\usepackage[english]{babel}
\usepackage{hyperref}
\usepackage[sort&compress,numbers]{natbib}
\usepackage{xcolor}
\usepackage{pslatex}
\usepackage{amsmath}
\usepackage{amssymb}
\usepackage{bbm}
\usepackage{euscript}
\usepackage{epsfig}
\usepackage{graphicx}
\usepackage{setspace}\definecolor{kkcolor}{rgb}{1,0,0}
\definecolor{kkcolor}{rgb}{1,0,0}

\newcommand\kkout{\marginpar{\color{kkcolor}$\clubsuit$}
\bgroup\markoverwith{\color{kkcolor}{\rule[04ex]{2pt}{0.8pt}}}\ULon}

\newcommand{\crd}{\color{black}}

\usepackage[colorinlistoftodos,prependcaption,textsize=tiny]{todonotes}

\newcommand{\mincas}{{\tt MINCAS}}

\begin{document}

\begin{titlepage}

\begin{center}
\end{center}
\begin{flushright}
\bf IFJPAN-IV-2020-7 \\
    September 2020
\end{flushright}

\vspace{5mm}
\begin{center}
    {\LARGE\bf 
     Medium induced QCD cascades: broadening and rescattering during branching} 
\end{center}

\vskip 5mm
\begin{center}
{\large E.\ Blanco$\,^a$, K.\ Kutak$\,^a$,
        W.\ P\l{}aczek$\,^b$,
        M.\ Rohrmoser$\,^a$ and R.\ Straka$\,^c$
        }
\vskip 2mm
{\em $^a\,$Institute of Nuclear Physics, Polish Academy of Sciences,\\
  ul.\ Radzikowskiego 152, 31-342 Krak\'ow, Poland}
\\
\vspace{1mm}
{\em $^b\,$Institute of Applied Computer Science, Jagiellonian University,\\
ul.\ \L{}ojasiewicza 11, 30-348 Krak\'ow, Poland}
\\
\vspace{1mm}
{\em $^c\,$ AGH University of Science and Technology, Krak\'ow, Poland}
\end{center}
 
\vspace{20mm}
\begin{abstract}
\noindent
We study evolution equations describing jet propagation through quark--gluon plasma (QGP). In particular we investigate the contribution of momentum transfer during branching  and find that such a contribution is sizeable. Furthermore, we study various approximations, such as the Gaussian approximation and the diffusive approximation to the jet-broadening term. We notice that in order to reproduce the BDIM equation (without the momentum transfer in the branching) the diffusive approximation requires a very large value of the jet-quenching parameter $\hat q$. 
\end{abstract}


\end{titlepage}

\section{Introduction}
\label{sec:Intro}

Quantum Chromodynamics (QCD) is the very well established theory of strong interactions with rich structure and many phases \cite{Ioffe:2010zz}.  
Here we want to focus on a jet-quenching phenomenon predicted in \cite{Gyulassy:1990ye,Wang:1991xy}, and observed experimentally at RHIC \cite{Adler:2002tq} and LHC  \cite{Aad:2010bu}. The jet quenching is a suppression of propagation of jets in quark--gluon plasma (QGP) due to jet--plasma interactions. This process has many phases, recently discussed in Refs.~\cite{Caucal:2018dla,Caucal:2019uvr}, see also \cite{Schlichting:2019abc}.
The jet-quenching phenomenon is approached from many directions: the kinetic theory \cite{Baier:2000mf,Baier:2000sb,Jeon:2003gi,Zakharov:1996fv,Zakharov:1997uu,Zakharov:1999zk,Baier:1994bd,Baier:1996vi,Arnold:2002ja,Ghiglieri:2015ala,Kurkela:2018vqr}, Monte Carlo methods \cite{Salgado:2003gb,Zapp:2008gi,Armesto:2009fj,Schenke:2009gb,Lokhtin:2011qq,Casalderrey-Solana:2014bpa}, the AdS/CFT \cite{Liu:2006ug}.  Furthermore, it is a multi-scale problem which, however, allows for factorisation in time. 
In particular, according to Refs.~\cite{Caucal:2018dla,Caucal:2019uvr}, in the first phase the jet propagates according to the vacuum-like parton shower with ordering in an angle, while in the next stage the coherence is broken and jet propagates through plasma  experiencing elastic scatterings and branching -- in this stage there are many soft radiations and wide-angle emissions. In the last stage, when jet leaves medium, again the vacuum-like emissions dictate its time evolution. 
In this paper, we focus on the second phase of the jet propagation through QGP 
In particular, we investigate what is the contribution of momentum transfer during branching to the broadening pattern of the jet. To address this problem,  we solve the equation proposed in~\cite{Blaizot:2013vha,Blaizot:2014rla} which is a generalised version of the equation solved by three of us in Ref.~\cite{Kutak:2018dim}\footnote{For other approach which addresses the transverse-momentum dependence but neglects the large-$x$ parton's spectrum see the relaxing harmonic approximation of Ref.~\cite{Andres:2020vxs}.}.
In this approach, QGP is modelled by static centres and the jet interacts with it weakly, jet propagating through plasma branches according to BDMPS-Z mechanism \cite{Baier:2000mf,Baier:2000sb,Jeon:2003gi,Zakharov:1996fv,Zakharov:1997uu,Zakharov:1999zk,Baier:1994bd,Baier:1996vi,Arnold:2002ja,Ghiglieri:2015ala} and gets broader due to elastic scattering with plasma.  

In Ref.~\cite{Kutak:2018dim} it has been observed that accounting for the broadening term beyond the diffusive approximation leads to the non-Gaussian broadening for jet observables. It turns out that the non-Gaussianity leads to much stronger broadening of the cross section for decorrelations than the Gaussian approximation.

In the current study, we propose detail study of impact of momentum exchange during branching and its contribution to the broadening. Experimentally, the broadening is rather small and its observation at the LHC energies is hindered by the vacuum effects \cite{Mueller:2016gko,Mueller:2016xoc}. There are also possible effects which could give negative contribution to the broadening \cite{Zakharov:2019fro,Zakharov:2020sfx,Zakharov:2020whb}. Furthermore, a more realistic model of the medium accounting for its expansion \cite{Iancu:2018trm,Zakharov:2020sfx,Adhya:2019qse} will probably reduce the amount of the broadening.
While at present we do not account for a more realistic scenario, i.e.\ the expansion of the medium we mimic it by scaling the $\hat q$ parameter. We observe that reducing its value leads to the smaller broadening.\\  
The paper is organised as follows:
In Section~2 we discuss the version of the transverse-momentum-dependent BDIM (Blaizot, Dominguez, Iancu, Mehtar-Tani) equation \cite{Blaizot:2014rla} where the momentum transfer in the kernel is taken into account and we present its solution with the use of Monte Carlo methods. In Section~3 we compare the BDIM equation to some of its approximations, i.e. the case where transverse momentum in the branching kernel is neglected, the case when the broadening term is represented by the diffusive approximation, and the Gaussian approximation where the transverse momentum and the longitudinal momentum are factorised.  
We conclude our work in Section~5.
In Appendix~A we present one of the Monte Carlo algorithms for solving
the full BDIM equation%
\footnote{The other one is an extension of the algorithm employed in
the Monte Carlo program \mincas, described in Ref.~\cite{Kutak:2018dim}, and will be presented elsewhere.}, 
while in Appendix~B we describe a numerical method
used to solve the diffusive approximation of the BDIM equation.

\section{Momentum-transfer-dependent BDIM equation and its solution}
\label{sec:md-BDIM}

The evolution equation for the gluon transverse-momentum-dependent
distribution $D(x,\mathbf{k},t)$ in the dense medium reads
\cite{Blaizot:2014rla}
\begin{equation}
\begin{aligned}
\frac{\partial}{\partial t} D(x,\mathbf{k},t) = & \:  \alpha_s \int_0^1 dz\, \int\frac{d^2q}{(2\pi)^2}\left[2{\cal K}(\mathbf{Q},z,\frac{x}{z}p_0^+) D\left(\frac{x}{z},\mathbf{q},t\right) 
- {\cal K}(\mathbf{q},z,xp_0^+)\, D(x,\mathbf{k},t) \right] \\
+& \int \frac{d^2\mathbf{l}}{(2\pi)^2} \,C(\mathbf{l})\, D(x,\mathbf{k}-\mathbf{l},t).
\end{aligned}
\label{eq:BDIM1}
\end{equation}
The kernel ${\cal K}(\mathbf{Q},z,xp_0^+)$ which accounts for the momentum-dependent medium induced branching is given by
\begin{equation}
{\cal K}(\mathbf{Q},z,p_0^+)=\frac{2}{p_0^+}\frac{P_{gg}(z)}{z(1-z)}\sin\left[\frac{\mathbf{Q}^2}{2k_{br}^2}\right]\exp\left[-\frac{\mathbf{Q}^2}{2k_{br}^2}\right] 
\label{eq:Kqz}
\end{equation}
with
\begin{equation}
\omega=xp_0^+,\,\,\,\, k_{br}^2=\sqrt{\omega_0\hat q_0},\,\,\,\,\,\mathbf{Q}=\mathbf{k}-z\,\mathbf{q},\,\,\, \omega_0=z(1-z)p_0^+ 
\end{equation}
and
\begin{equation}
\,\,\hat q_0=\hat q f(z),\,\, f(z)=1-z(1-z),\,\,\, P_{gg}(z)=N_c\frac{\left[1-z(1-z)\right]^2}{z(1-z)}.    
\end{equation}
where $p_0^+\equiv E$ is energy of jet entering the medium,
$x$ -- is longitudinal momentum fraction of mini jet, 
$\mathbf{k}=(k_x,k_y)$ -- is transverse-momentum vector of mini jet,
$\hat{q}$ -- the quenching parameter,
$\alpha_s$ -- the QCD coupling constant
and $N_c$ -- the number of colours.

The elastic collision kernel $C(\mathbf{l})$ is given by
\begin{equation}
C(\mathbf{l}) = w(\mathbf{l}) - \delta(\mathbf{l}) \int d^2\mathbf{l'}\,w(\mathbf{l'})\,,
\label{eq:Cq}
\end{equation}
where the function $w(\mathbf{l})$ 
models the momentum distribution of medium quasi-particles.
We consider two scenarios:
\begin{enumerate}
    \item The out-of-equilibrium distribution~\cite{Blaizot:2014rla}:
\begin{equation}
 w(\mathbf{l}) = \frac{16\pi^2\alpha_s^2N_cn}{\mathbf{l}^4}\,,
\label{eq:wq1}
\end{equation}
with $\mathbf{l}=(l_x,l_y)$ being transverse-momentum vector 
and $n$ -- the density of scatterers.
\item The situation where the medium  equilibrates and the transverse-momentum distribution assumes the form obtained from the Hard Thermal Loops (HTL) calculation \cite{Aurenche:2002pd}. In this case the medium is characterised by a mass scale given by the Debye mass $m_D$:
\begin{equation}
 w(\mathbf{l}) = \frac{g^2m_D^2T}{\mathbf{l}^2(\mathbf{l}^2+m_D^2)}\,,
\label{eq:wq2}
\end{equation}
$$
m_D^2=g^2T^2\left(\frac{N_c}{3}+\frac{N_f}{6}\right),\quad
g^2 = 4\pi\alpha_s.
$$
\end{enumerate}

The equation (\ref{eq:BDIM1}) has been solved using the Monte Carlo program \mincas\ by extending the algorithm presented in \cite{Kutak:2018dim} (to be described elsewhere) and, independently, using another Monte Carlo algorithm described in the Appendix~A. 
The two solutions have been checked to be in a good numerical agreement.
Here we present the results from \mincas\ obtained using the following input parameters:
\begin{center}
\begin{tabular}{ c c c }
 $x_{\text{min}}=10^{-4}$,  & $\epsilon=10^{-6}$ & \\
 $q_{\text{min}}=0.1\,$GeV,  & $m_D=0.993\,$GeV, & $\sigma_{k_0}=0\,$GeV, \\
 $N_c=3$, & $\overline{\alpha}_s=0.3$, & \\
$E=100\,$GeV, & $n=0.243\,$GeV$^3$, & $\hat{q}= 1\,$GeV$^2/fm$.
\end{tabular}
\end{center}

\begin{figure}[!ht]
\centering{}
\includegraphics[scale=0.39]{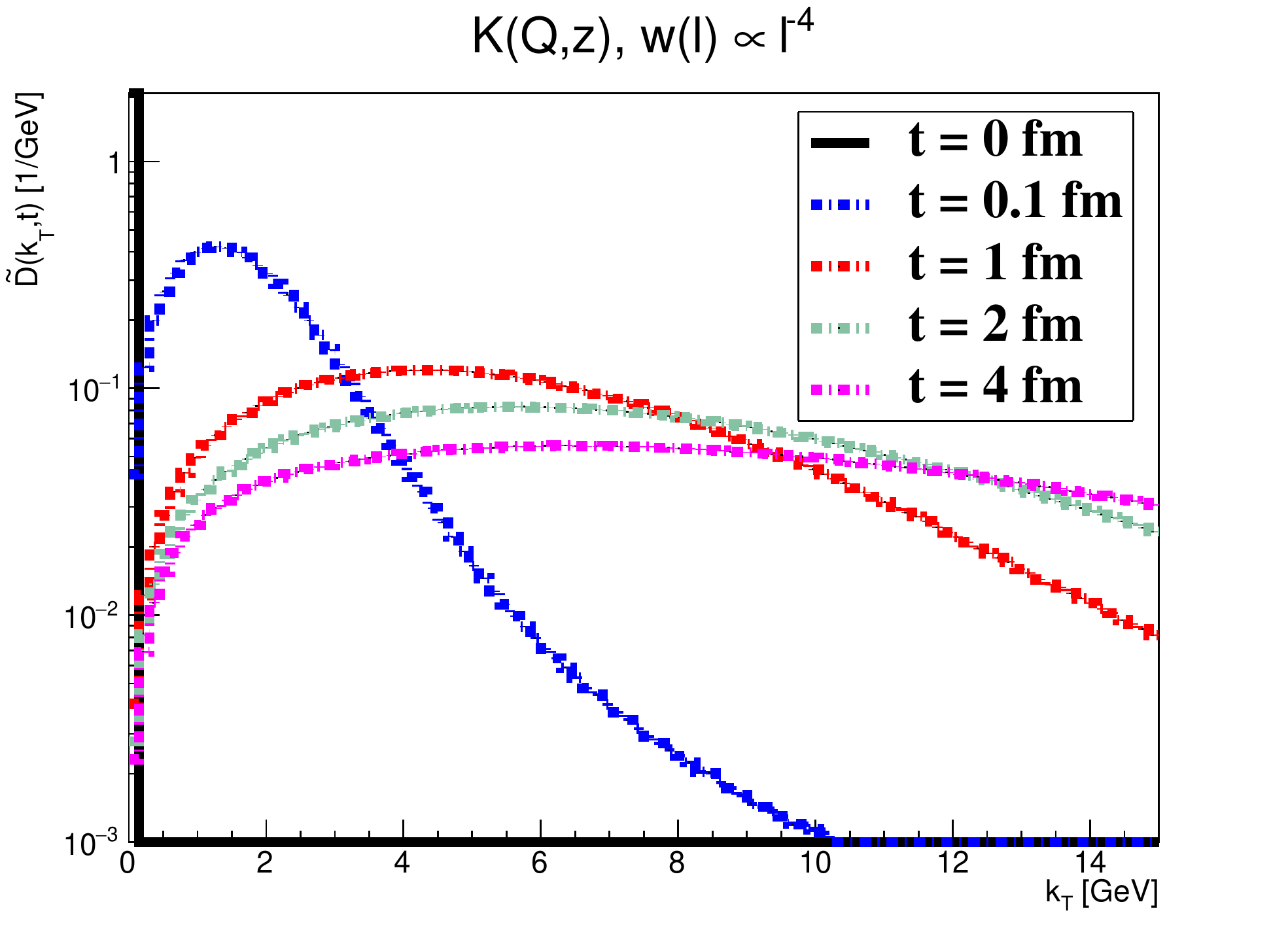}
\includegraphics[scale=0.39]{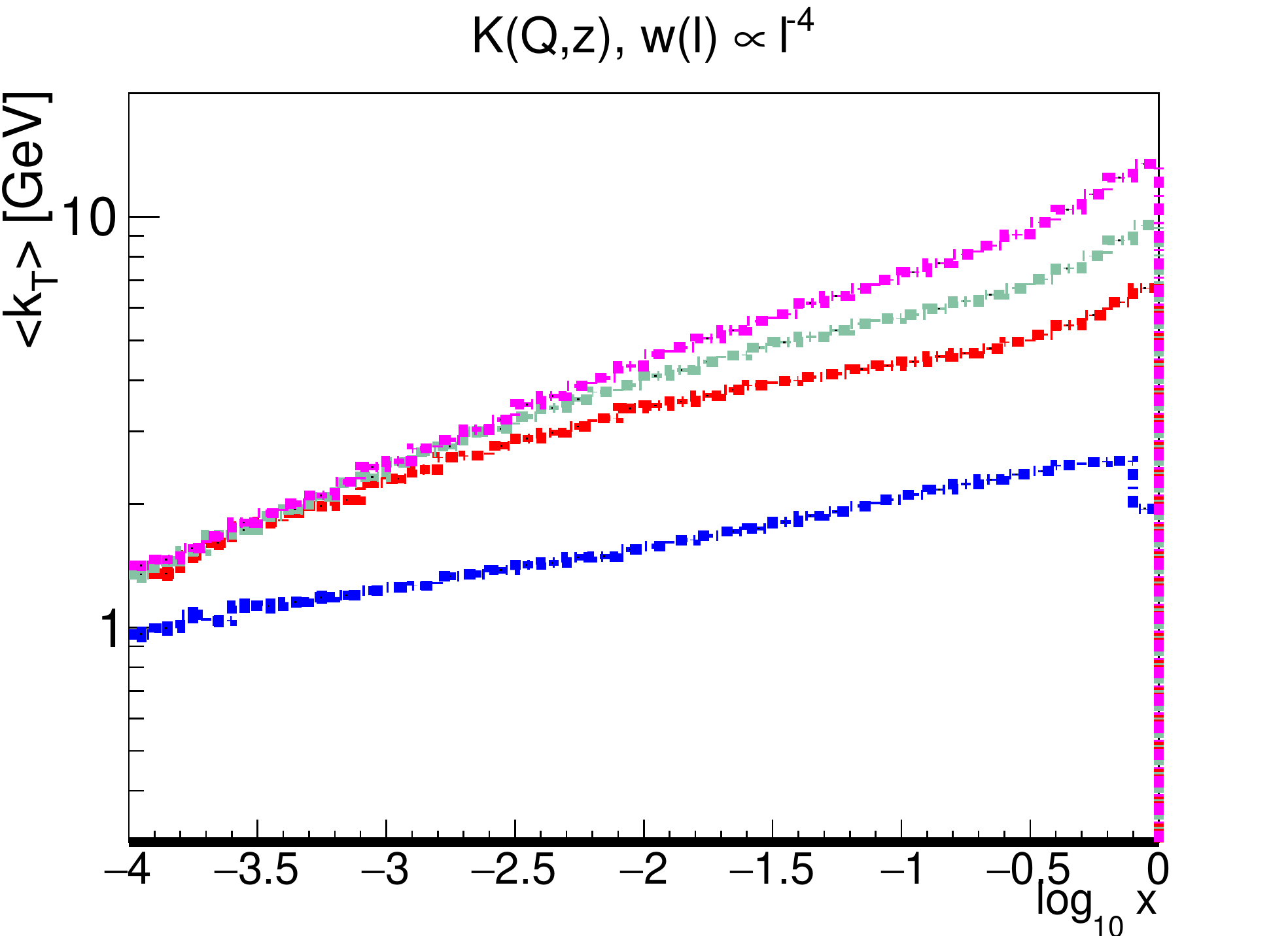}
\caption{The $k_T$ and $\langle k_T \rangle$ vs.\ $\log_{10}x$ distributions for the evolution time values $t=0, 0.1, 1, 2, 4\,$fm, for the full {\crd emission} kernel 
{\crd ${\cal K}(\mathbf{Q},z,p^+)$ (denoted as $\rm K(Q,z)$)} 
and the collision term of Eq.~(\ref{eq:wq1}).}
\label{fig:sol1}
\end{figure}

In Fig.~\ref{fig:sol1} we show the $k_T$ distributions as well as $\langle k_T\rangle$ as a function of $x$ for the evolution time values $t=0, 0.1, 1, 2, 4\,$fm. The detailed discussion of the solution is presented in the next section where we also discuss comparisons to the approximations of the BDIM equation.

\section{Comparison of BDIM equation to its approximations}
\label{sec:comp}

In this section we will discuss various approximations of the BDIM equation. 
\begin{itemize}
\item
The first approximation that we consider is the case when the momentum transfer during branching is neglected. In this case, as demonstrated in Ref.~\cite{Blaizot:2013vha}, the branching kernel simplifies to a purely collinear one and the transverse momentum dependence comes basically from the elastic scattering. The equation reads
    \begin{equation}
\begin{aligned}
\frac{\partial}{\partial t} D(x,\mathbf{k},t) = & \: \frac{1}{t^*} \int_0^1 dz\, {\cal K}(z) \left[\frac{1}{z^2}\sqrt{\frac{z}{x}}\, D\left(\frac{x}{z},\frac{\mathbf{k}}{z},t\right)\theta(z-x) 
- \frac{z}{\sqrt{x}}\, D(x,\mathbf{k},t) \right] \\
+& \int \frac{d^2\mathbf{q}}{(2\pi)^2} \,C(\mathbf{q})\, D(x,\mathbf{k}-\mathbf{q},t),
\end{aligned}
\label{eq:BDIM2}
\end{equation}
 where 
    \begin{equation}
       {\cal K}(z)= \frac{(1-z+z^2)^{5/2}}{[z(1-z)]^{3/2}}, \,\,\,\,\frac{1}{t^\ast}=\frac{\alpha_s N_c}{\pi}\sqrt{\frac{\hat{q}}{p_0^+}}\,.
    \end{equation}
\item
One can further simplify the BDIM equation by expanding the elastic collision term and using the diffusive approximation \cite{Blaizot:2013vha} to obtain
\begin{equation}
\begin{aligned}
\frac{\partial}{\partial t} D(x,\mathbf{k},t) = & \: \frac{1}{t^*} \int_0^1 dz\, {\cal K}(z) \left[\frac{1}{z^2}\sqrt{\frac{z}{x}}\, D\left(\frac{x}{z},\frac{\mathbf{k}}{z},t\right)\theta(z-x) 
- \frac{z}{\sqrt{x}}\, D(x,\mathbf{k},t) \right] \\
+&\frac{1}{4}\hat q \nabla^2_k\bigg[D(x,\mathbf{k},t)\bigg].
\end{aligned}
\label{eq:ktee1_diff}
\end{equation}
In the above equation, as compared to Ref.~\cite{Blaizot:2013vha}, we have neglected the mild logarithmic dependence of $\hat q$ in the diffusion term on $k_T$.
\item 
Eq.~(\ref{eq:ktee1_diff}) was also solved approximately in Ref.~\cite{Blaizot:2014ula}.  To arrive at the solution, the branching term was neglected and the Gaussian ansatz was used. The solution reads
\begin{equation}
D(x,\mathbf{k},t) = D(x,t)\,\frac{4\pi}{\langle k_\perp^2\rangle}
\exp\left[-\frac{\mathbf{k}^2}{\langle k_\perp^2\rangle}\right],
\end{equation}
where 
\begin{equation}
\langle k_\perp^2\rangle=\min\left\{\frac{1}{2}\hat q t(1+x^2),\, \frac{k^2_{br}(x)}{4\bar\alpha},
\,(x E)^2\right\}, \quad k_{br}^2(x)=\sqrt{x E \hat q}.   
\end{equation}
In the above, it is assumed that $k_\perp^2<\omega^2=(xE)^2$, and the parameters are: $\bar\alpha_s=0.3$, $\hat q=1\,$GeV$^2/$fm, $E=100\,$GeV.

\end{itemize}

Similarly to Eq.~(\ref{eq:BDIM1}), Eq.~(\ref{eq:BDIM2}) has been solved using the Monte Carlo programs, basically re-obtaining the result from Ref.~\cite{Kutak:2018dim},
while Eq.~(\ref{eq:ktee1_diff}) has been solved with the help of the numerical method described in Appendix~B.
For Eq.~(\ref{eq:BDIM1}) and Eq. (\ref{eq:BDIM2}) we have used both the functions $w(\mathbf{l})$ from Eqs.~(\ref{eq:wq1}) and (\ref{eq:wq2}) to describe the collision term when using both the full kernel and the simplified one. 
In the case of the full kernel, we have also performed calculations without the collision term. For all the  the presented results, we have used the parameters given in the previous section. 

\begin{figure}[!ht]
\centering{}
\includegraphics[width=0.32\textwidth]{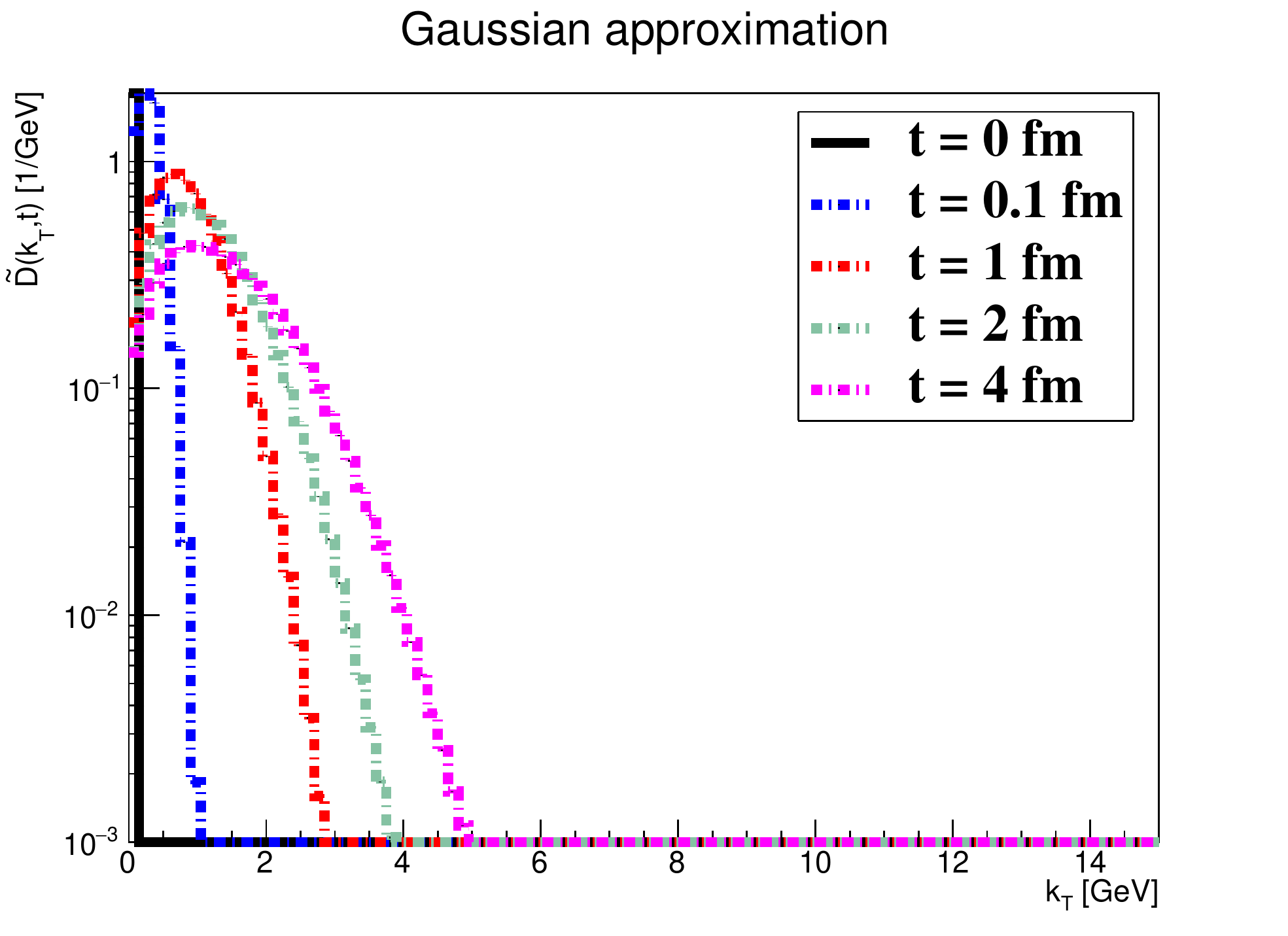}
\includegraphics[width=0.32\textwidth]{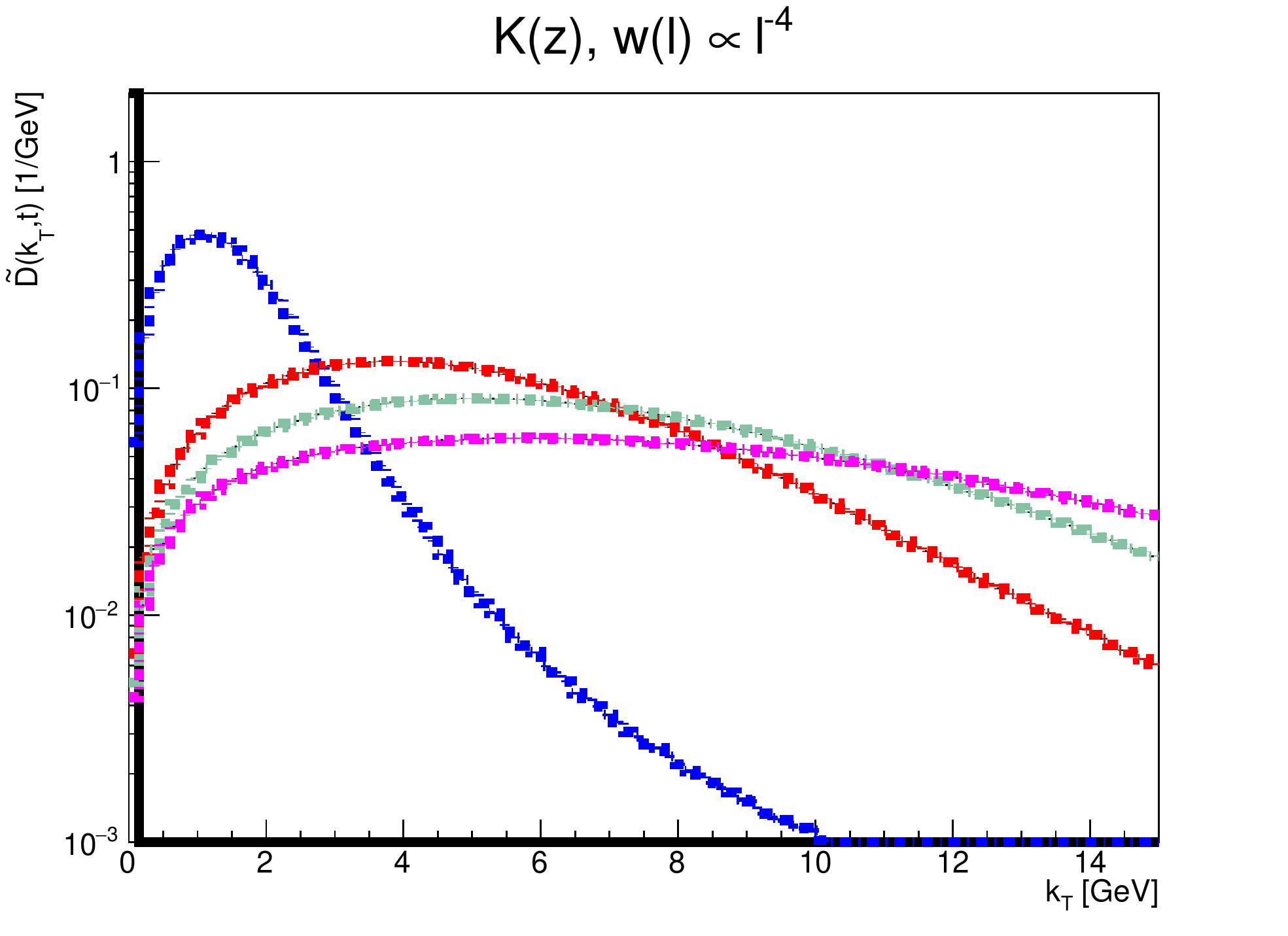}
\includegraphics[width=0.32\textwidth]{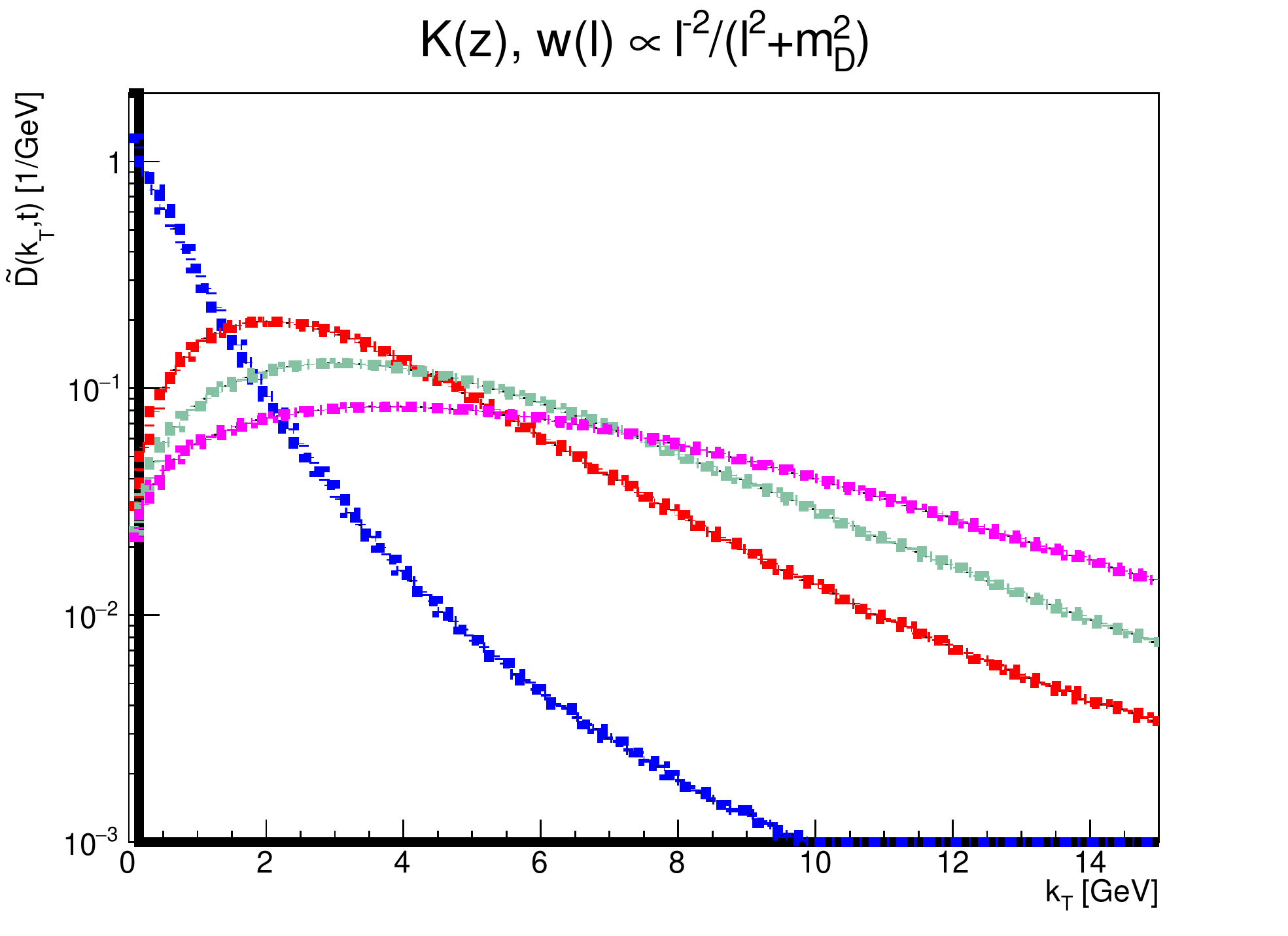}
\includegraphics[width=0.32\textwidth]{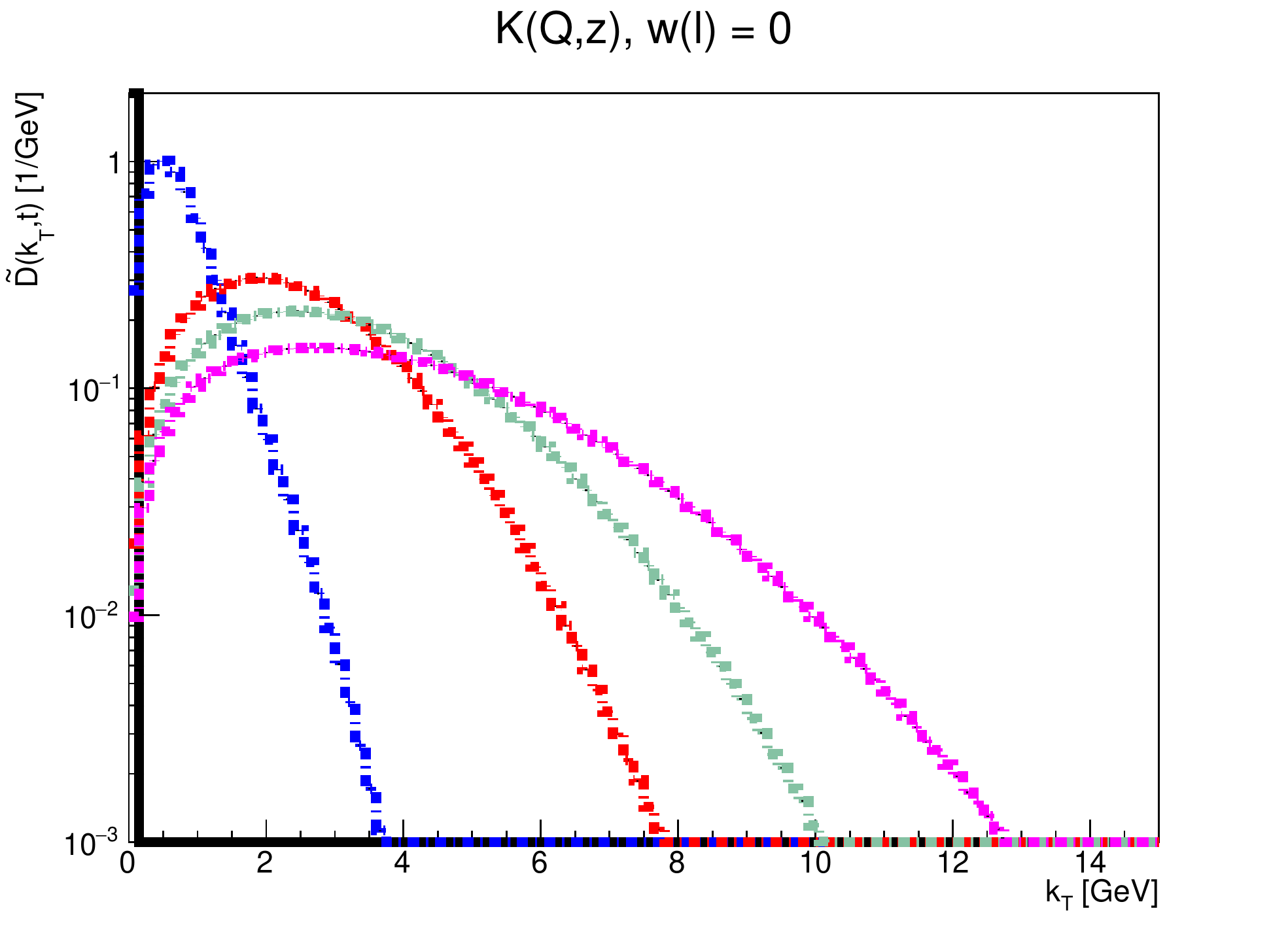}
\includegraphics[width=0.32\textwidth]{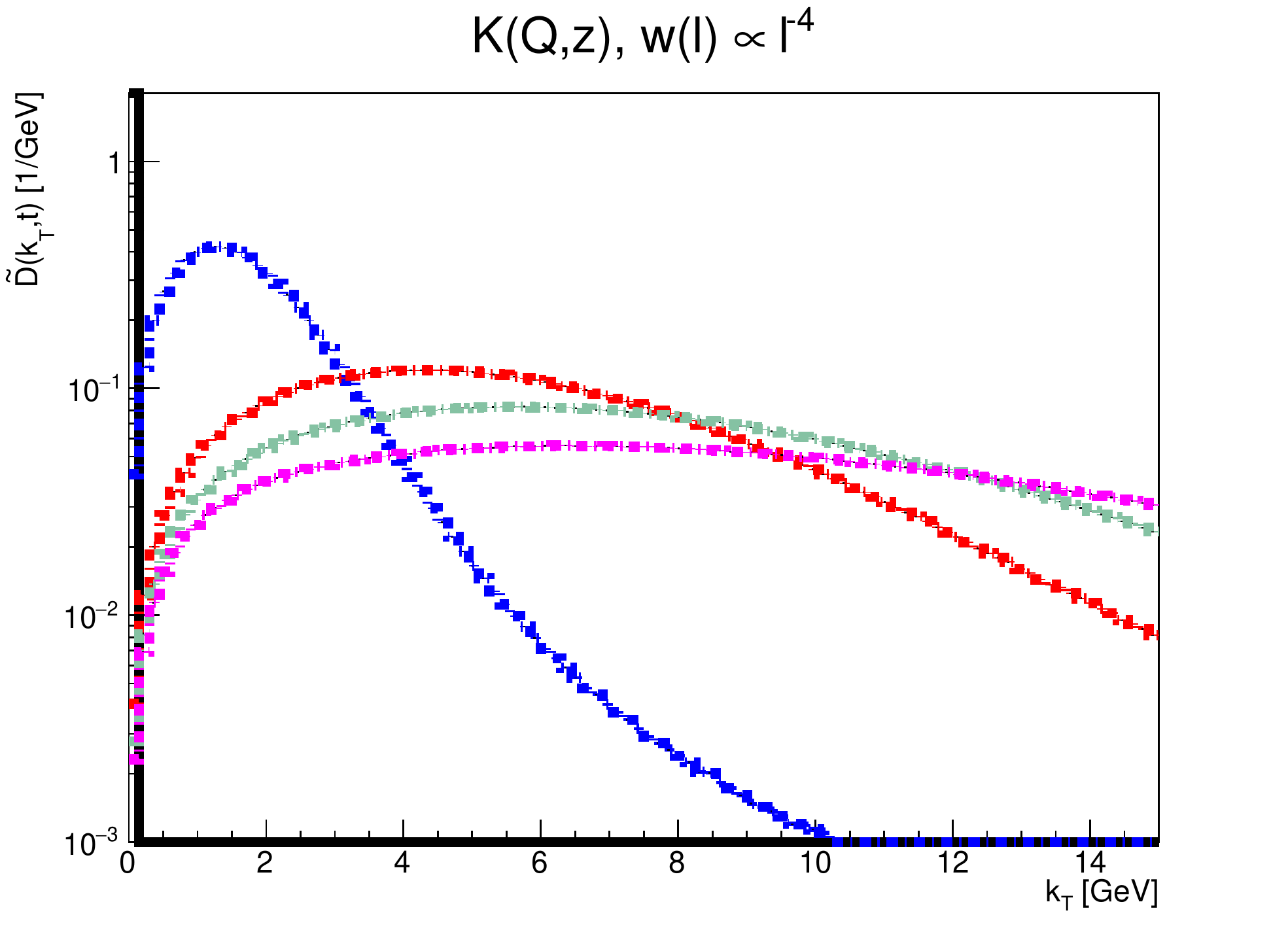}
\includegraphics[width=0.32\textwidth]{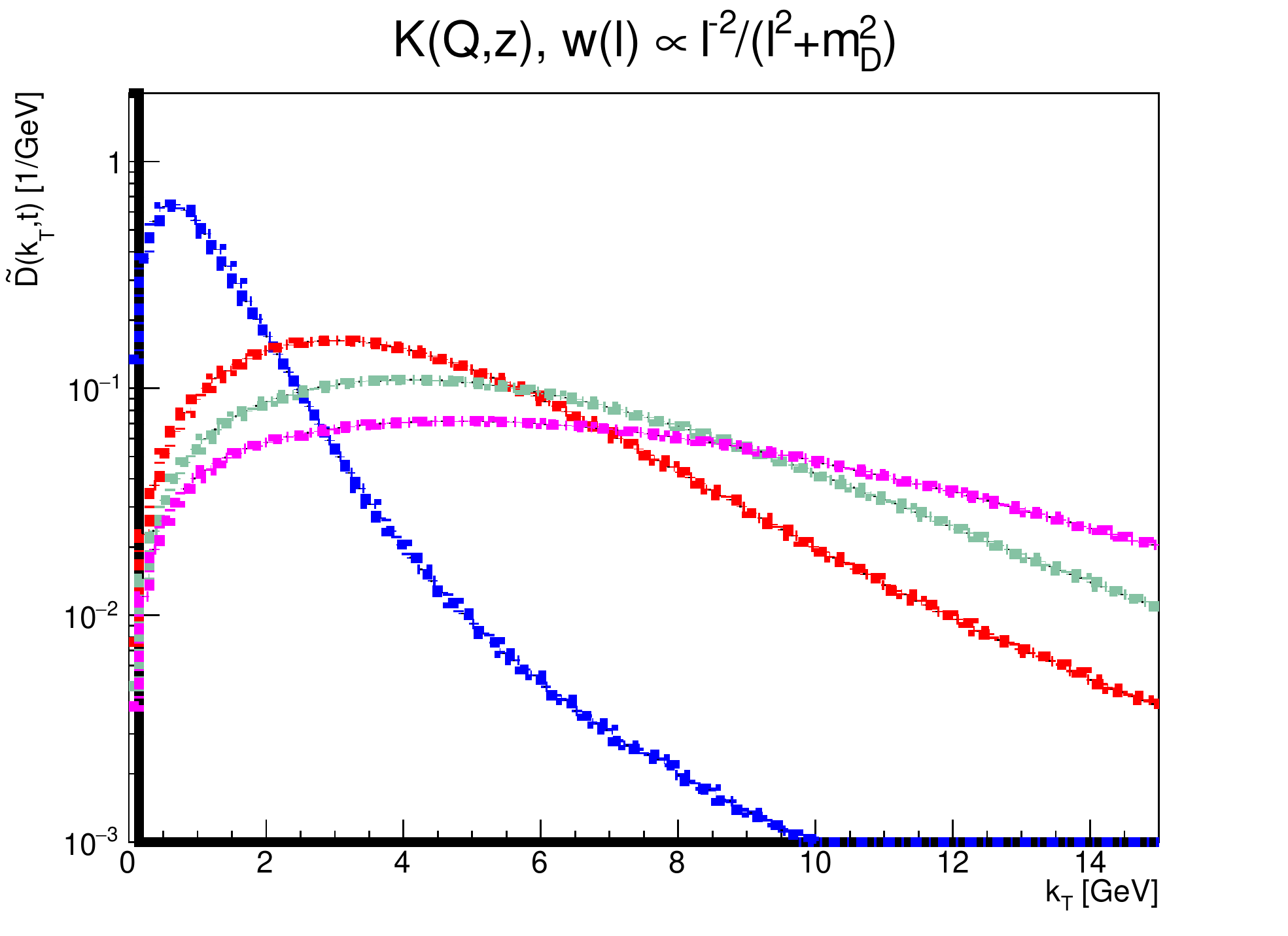}
\caption{The $k_T$ distributions for the evolution time values $t=0, 0.1, 1, 2, 4\,$fm, for different kernels: the Gaussian approximation, ${\cal K}(z)$ and {\crd ${\cal K}(\mathbf{Q},z,p^+)$ (denoted as $\rm K(Q,z)$}), and different collision terms: no collision term, the collision term as in Eq.~(\ref{eq:wq1}) and as in Eq.~(\ref{eq:wq2}).}
\label{fig:kTplots}
\end{figure}

In Fig.~\ref{fig:kTplots} 
we show the $k_T$ distributions of the six cases studied for evolution time values $t=0, 0.1, 1, 2, 4$ fm. We directly notice that the Gaussian approximation fails to describe any of the other results. The nearest distribution is the one with the full kernel and no collision term, which approaches a Gaussian shape, but with a much wider width. The other distributions (with the collision term) show fast broadening of the initial Dirac-$\delta$-like distribution, exhibiting the non-Gaussian shape. The broadening is faster with $w(\mathbf{l})$ given by Eq.~(\ref{eq:wq1}) than with the one given by Eq.~(\ref{eq:wq2}), i.e the broadening is faster with out-of-equilibrium momentum distributions of the medium quasi-particles.

\begin{figure}[!ht]
\centering{}
\includegraphics[width=0.32\textwidth]{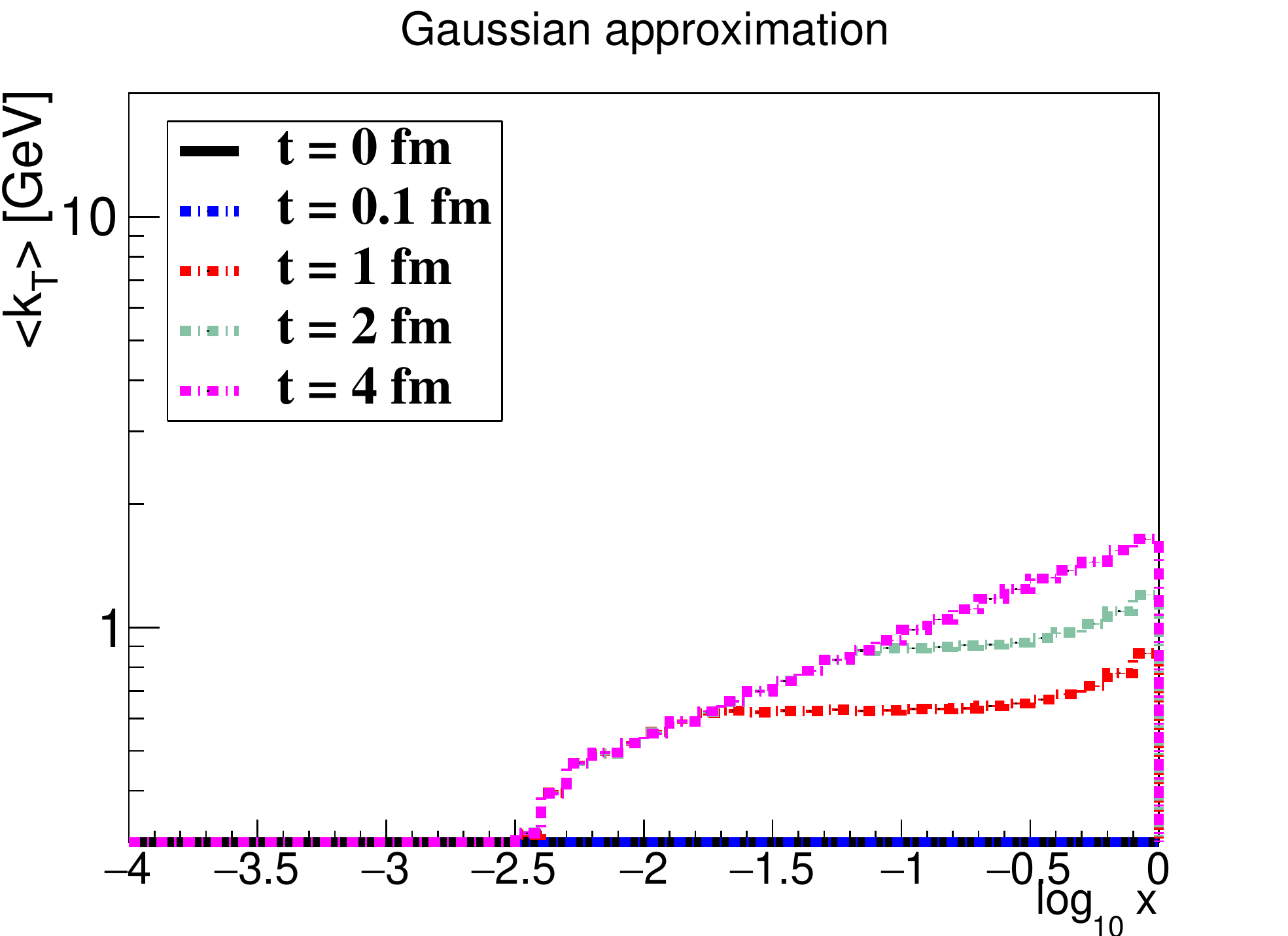}
\includegraphics[width=0.32\textwidth]{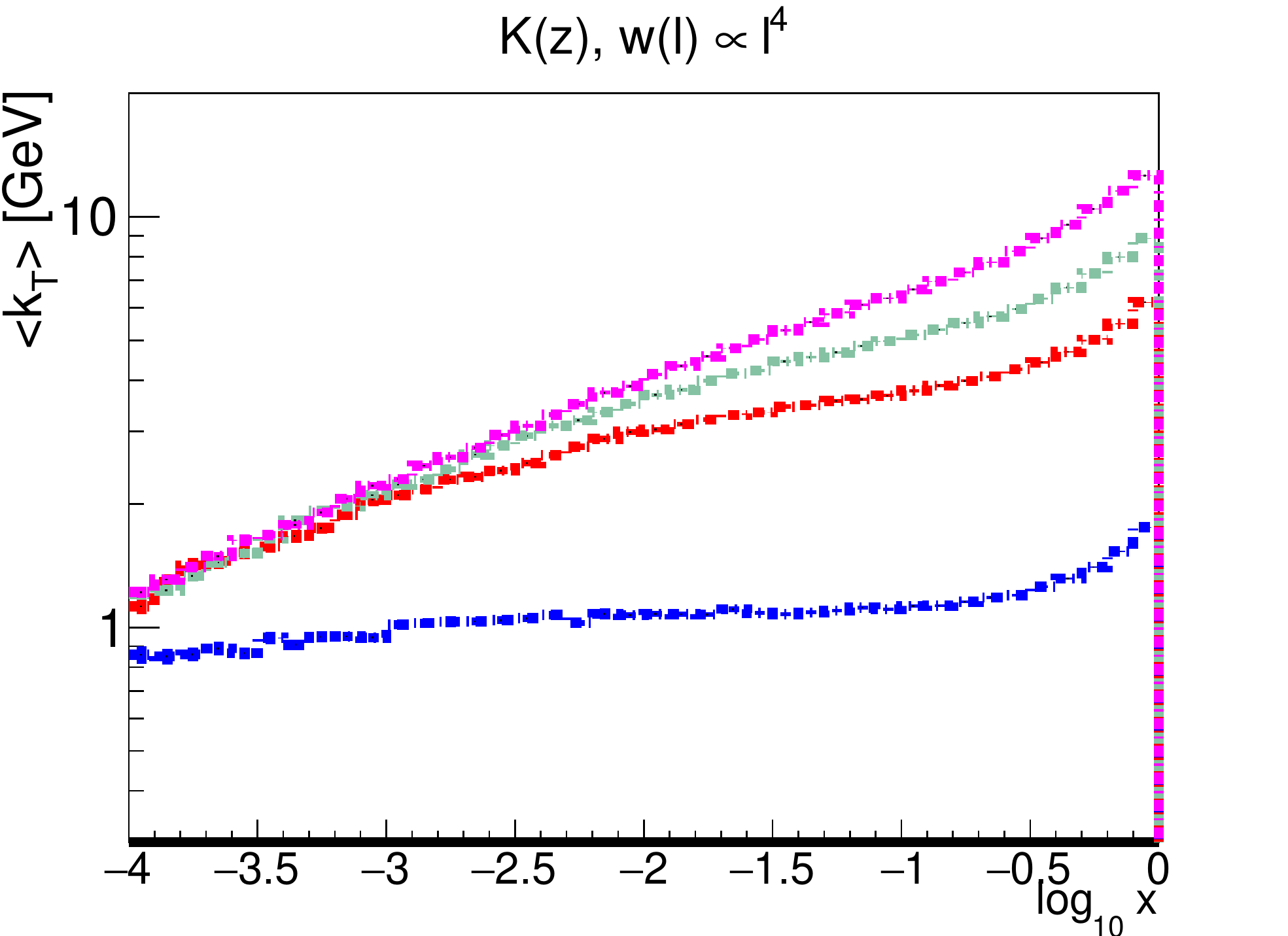}
\includegraphics[width=0.32\textwidth]{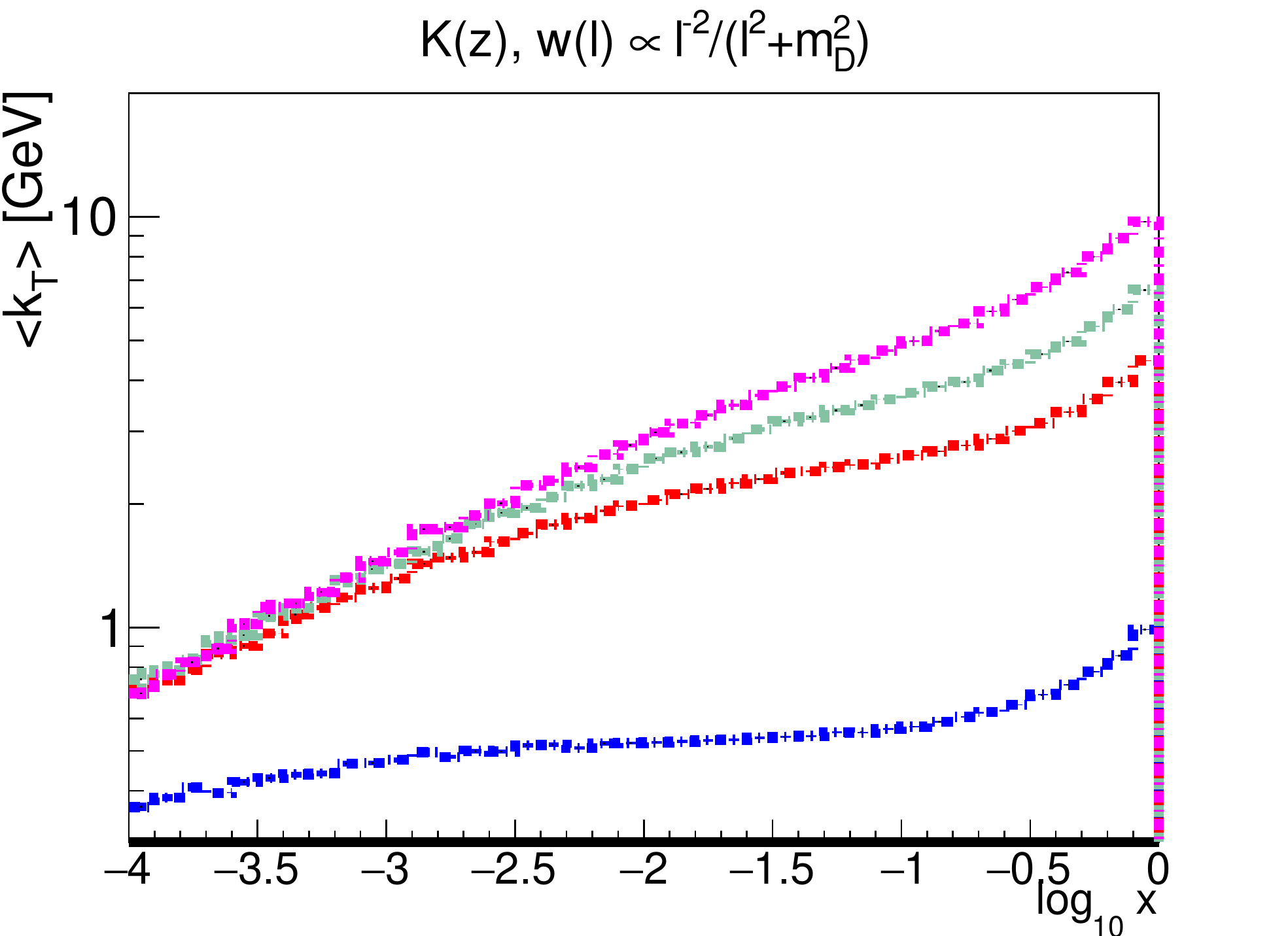}
\includegraphics[width=0.32\textwidth]{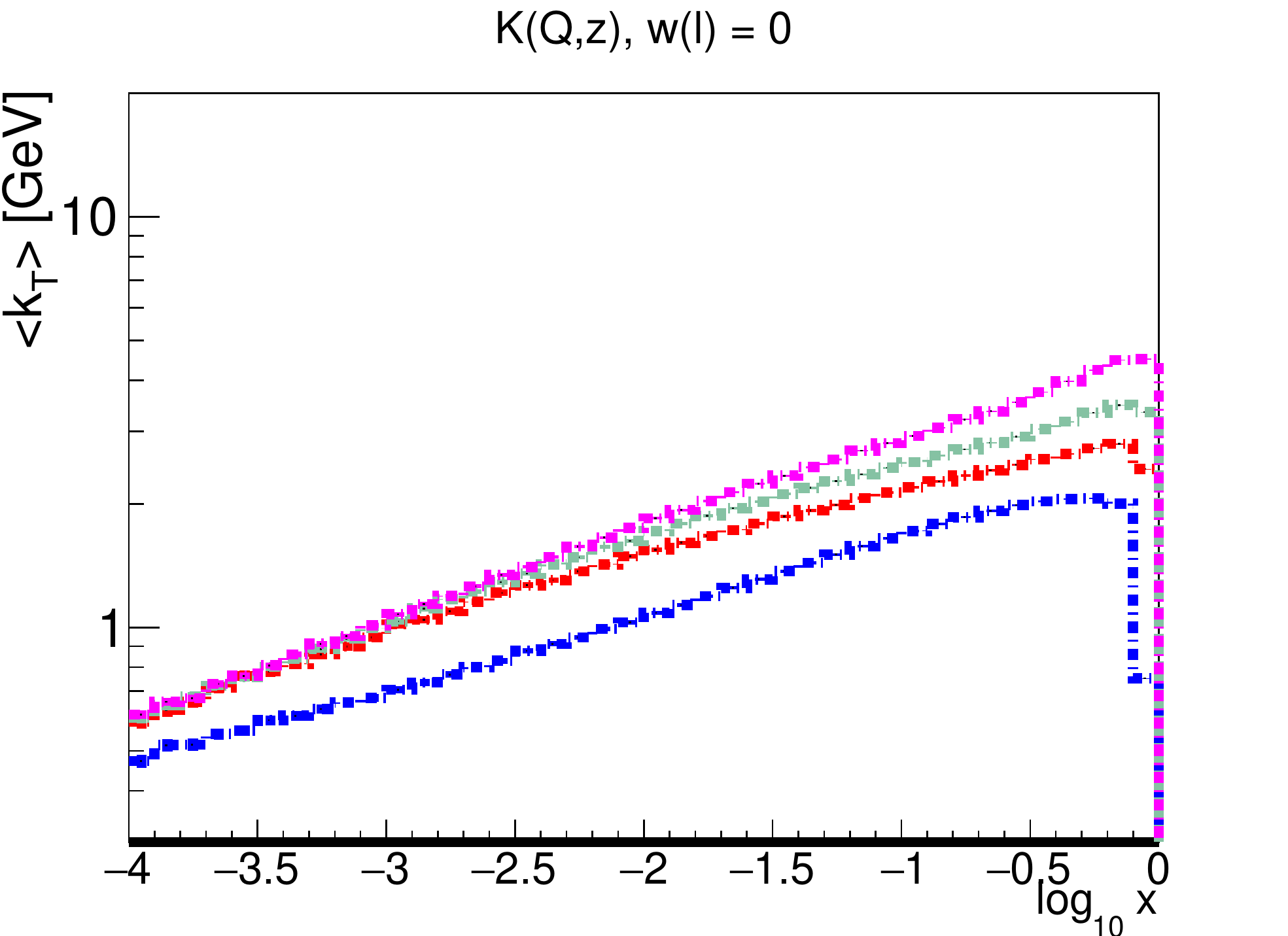}
\includegraphics[width=0.32\textwidth]{plots/mkTlogx_KzQ_w1_q=1.pdf}
\includegraphics[width=0.32\textwidth]{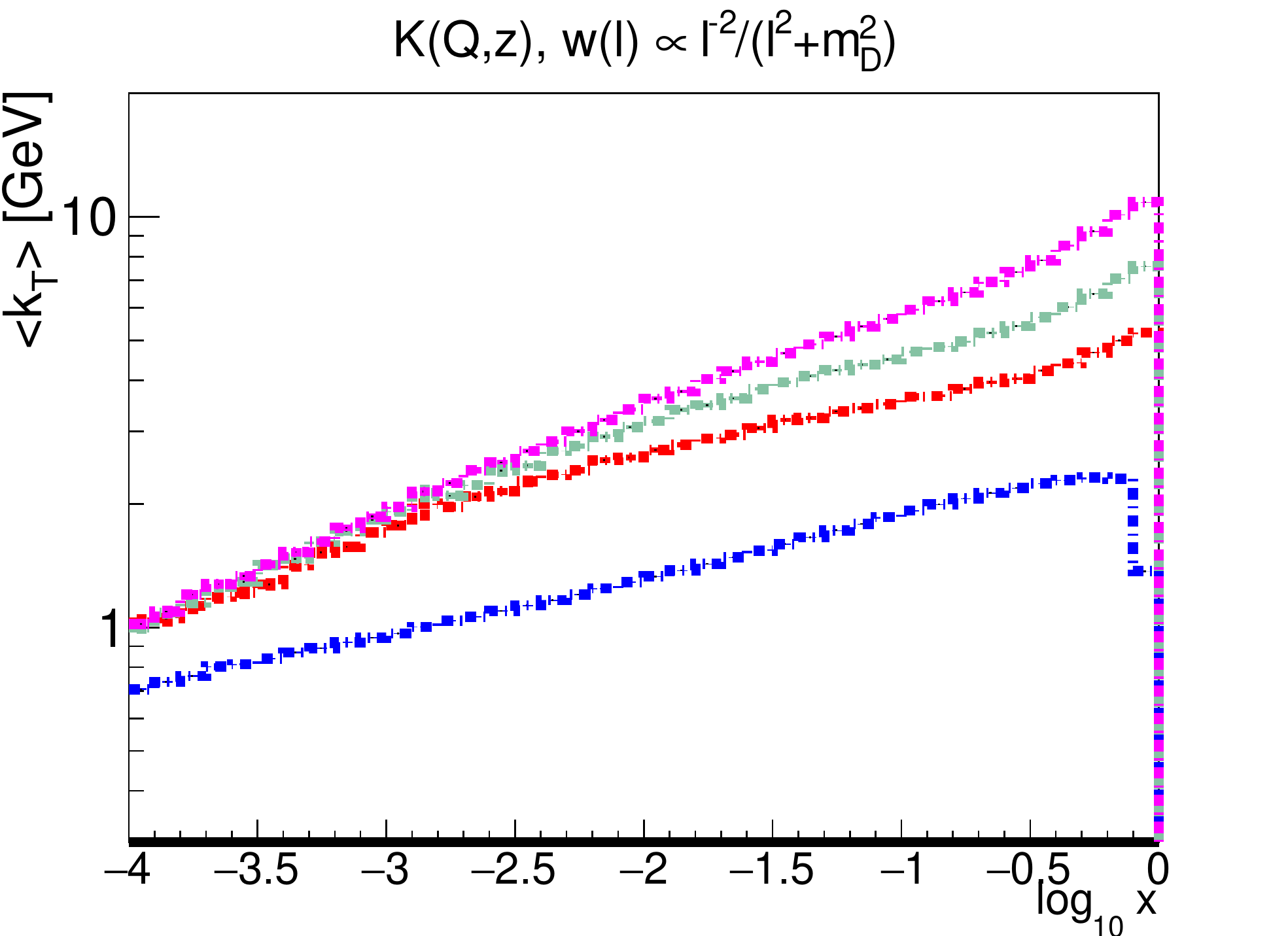}
\caption{The $\langle k_T \rangle$ vs.\ $\log_{10}x$ distributions for the evolution time values $t=0, 0.1, 1, 2, 4\,$fm, for different kernels: the Gaussian approximation, ${\cal K}(z)$ and {\crd ${\cal K}(\mathbf{Q},z,p^+)$ (denoted as $\rm K(Q,z)$)}, and different collision terms: no collision term, the collision term as in Eq. ~(\ref{eq:wq1}) and in Eq.~(\ref{eq:wq2}), respectively.}
\label{fig:mkxlogx_distr}
\end{figure}

In Fig.~\ref{fig:mkxlogx_distr} we present the dependence of the mean value of $k_T$ on $\log_{10} x$. For all cases, $\langle k_T \rangle$ grows with time and with $x$. It is still true for the Gaussian approximation, even if the distribution for the different evolution time join each other under certain values of $x$. We can clearly see in these figures a different behaviour around $x=1$ between the distributions corresponding to the $z$-only dependent kernel and the ones corresponding to the full kernel which show a drop.
{\crd 
This drop results from the fact that the evolution starts at $x=1$ with $k_T = 0$ 
and already a single soft emission with the $\mathbf{Q}$-dependent kernel $\mathcal{K}$
gives to the emitter a significant $k_T$-kick, 
which is not the case for the $z$-only dependent emission kernel. 
This effect is more pronounced for the shortest evolution time $t=0.1\,$fm, 
because in this case the  $(x,k_T)$-distribution is strongly peaked at $x=1$ and $k_T=0$, while for the longer evolution times this peak is smeared out, 
so the contribution from $x=1$ and $k_T=0$ to $\langle k_T \rangle$ is much smaller.
Except for the drop near $x=1$ for short evolution times with the full emission kernel, 
the $\langle k_T \rangle$ distributions increase  with $x$. 
The Gaussian approximation gives the lowest $\langle k_T \rangle$ values, while they are the highest for the evolution with the full emission kernel -- and these, in particular, are higher than in the case with the $z$-only dependent emission kernel.
This results from the fact that in the former case the $k_T$-broadening is produced 
not only in the collisions with the medium 
(due to the $C(\mathbf{l})$ term in Eq.~(\ref{eq:BDIM1})), but also in the emission process
(due to the $\mathbf{Q}$-dependence of the kernel $\mathcal{K}$ in Eq.~(\ref{eq:BDIM1})).
}

\begin{figure}[!ht]
\centering{}
\includegraphics[width=0.49\textwidth]{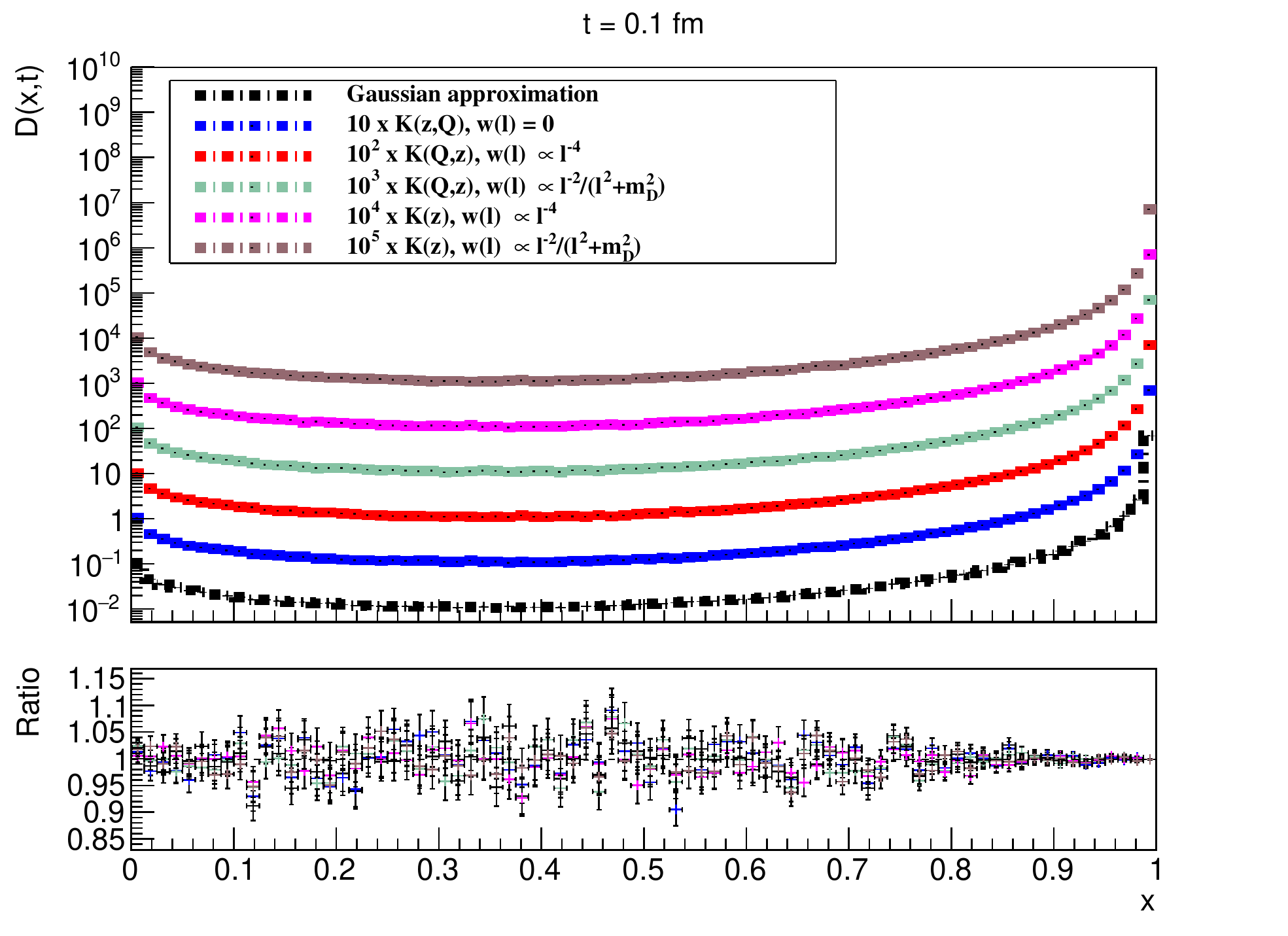}
\includegraphics[width=0.49\textwidth]{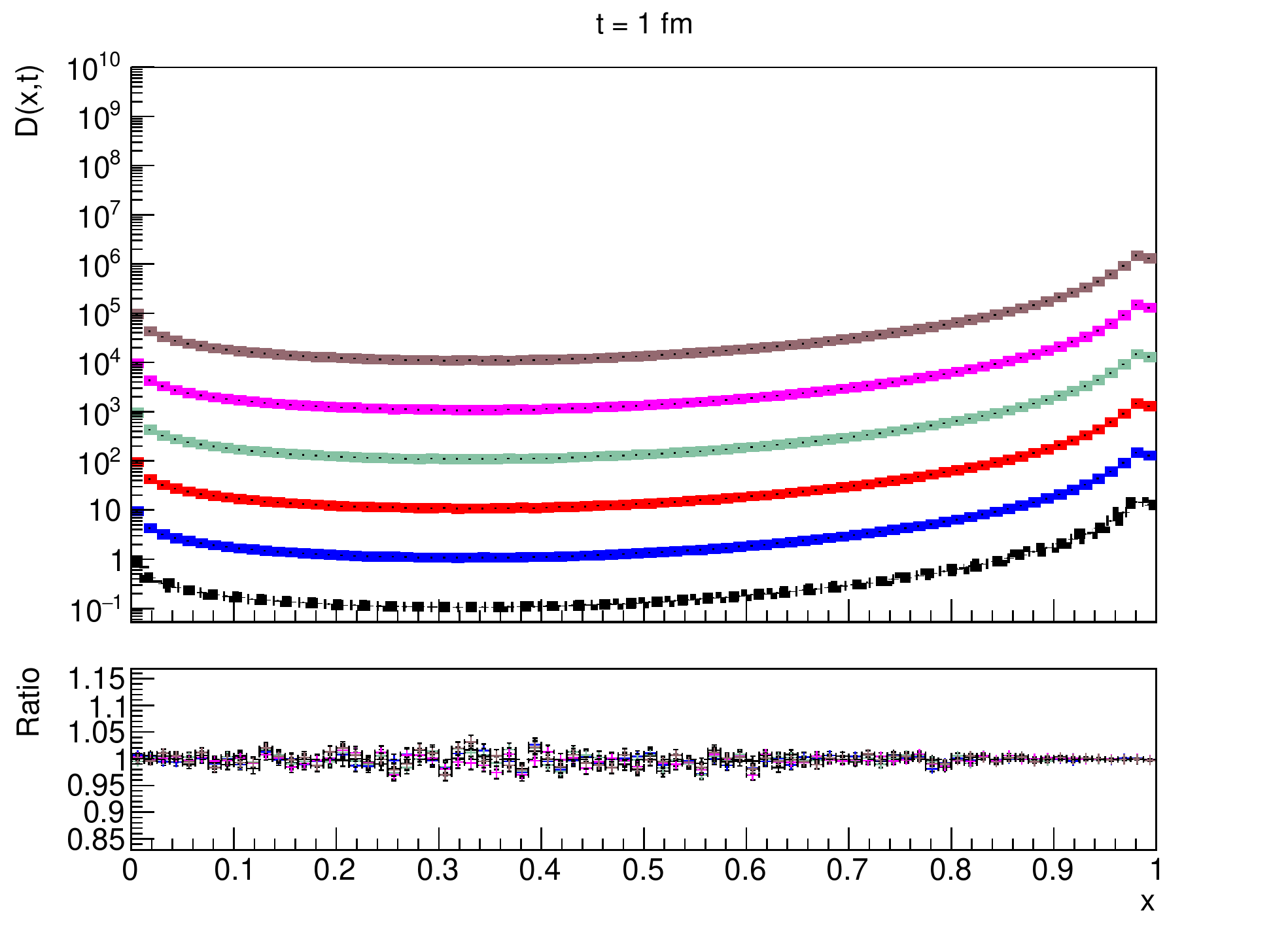}
\includegraphics[width=0.49\textwidth]{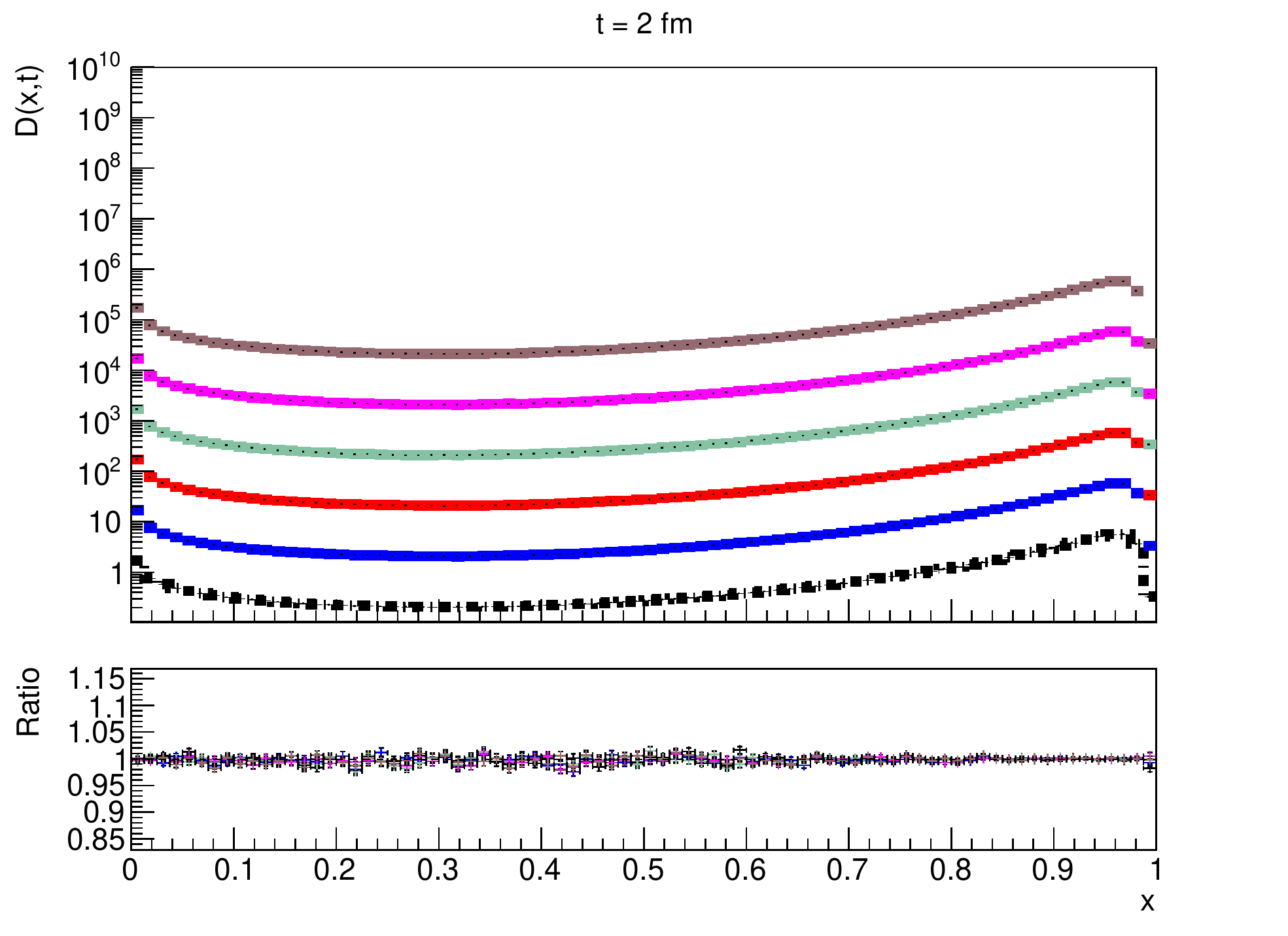}
\includegraphics[width=0.49\textwidth]{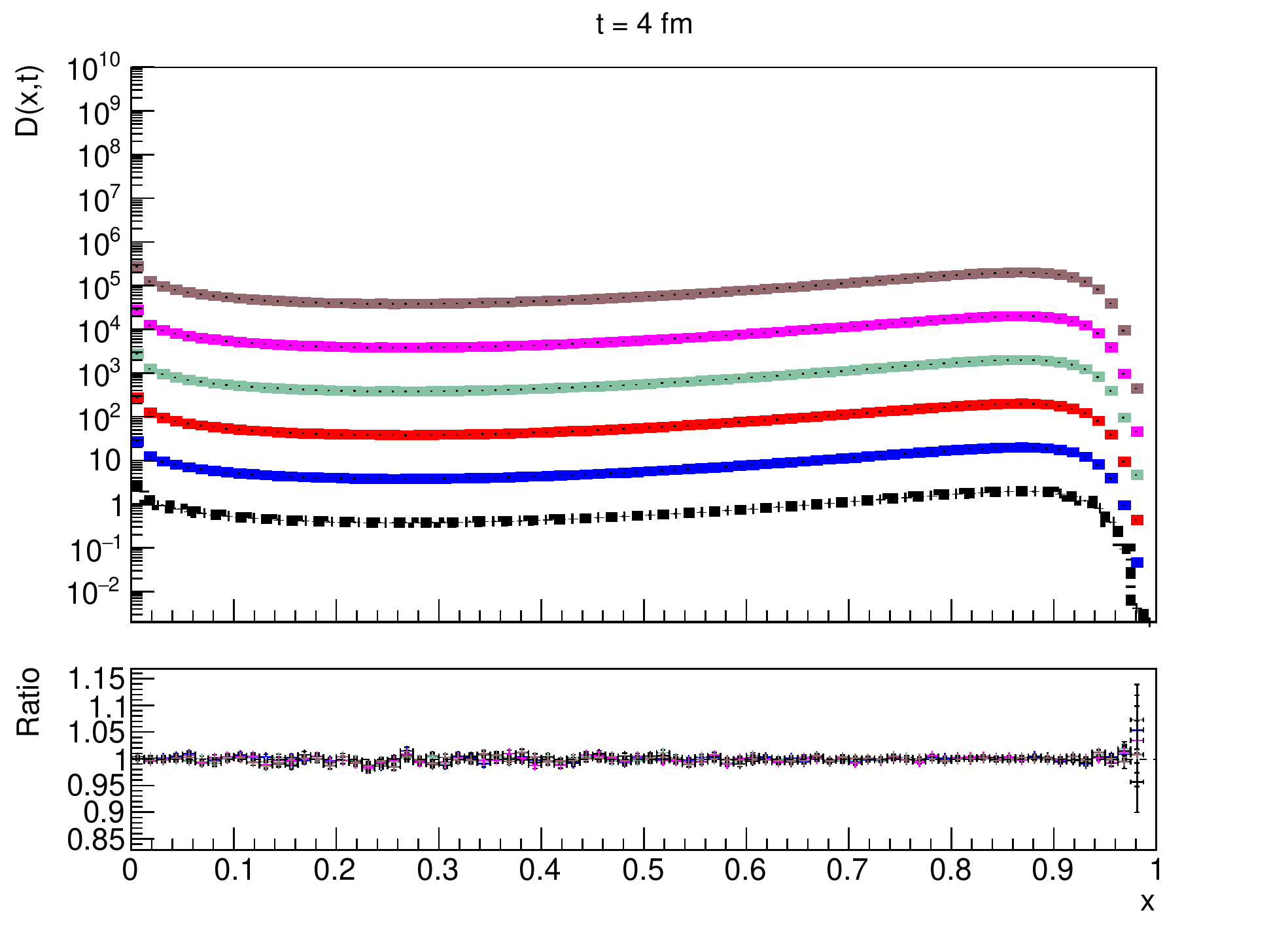}
\caption{The integrated $x$ distributions (multiplied by the factors $10^n,\, n=0,\ldots,5$) for the evolution time values $t=0.1, 1, 2, 4\,$fm, for different kernels: the Gaussian approximation, ${\cal K}(z)$ and ${\cal K}(\mathbf{Q},z)$, and different collision terms: no collision term, the collision term as in Eq.~(\ref{eq:wq1}) and as in Eq.~(\ref{eq:wq2}). The reference distribution used for the ratio plots is the one for the full kernel ${\cal K}(\mathbf{Q},z)$ and the collision term of Eq.~(\ref{eq:wq1}).}
\label{fig:x_distr_comp}
\end{figure}

In Fig.~\ref{fig:x_distr_comp} we present distributions integrated over the transverse momenta for four values of the evolution time: $t=0.1, 1, 2, 4\,$fm. We see that all the transverse-momentum-dependent distributions, as a consequence of momentum conservation, collapse to the same $x$-dependent distributions. This further confirms that the study of the transverse-momentum dependence allows for more detailed study of the dynamics of the branching process. 
It also constitutes an important numerical cross-check that all our algorithms 
for the transverse-momentum-dependent evolution satisfy the condition:
\begin{equation}
D(x,t) = \int d^2\mathbf{k}\; D(x,\mathbf{k},t)\,.
\label{eq:intoverk}
\end{equation}

\begin{figure}[!ht]
\centering{}
\includegraphics[width=0.49\textwidth]{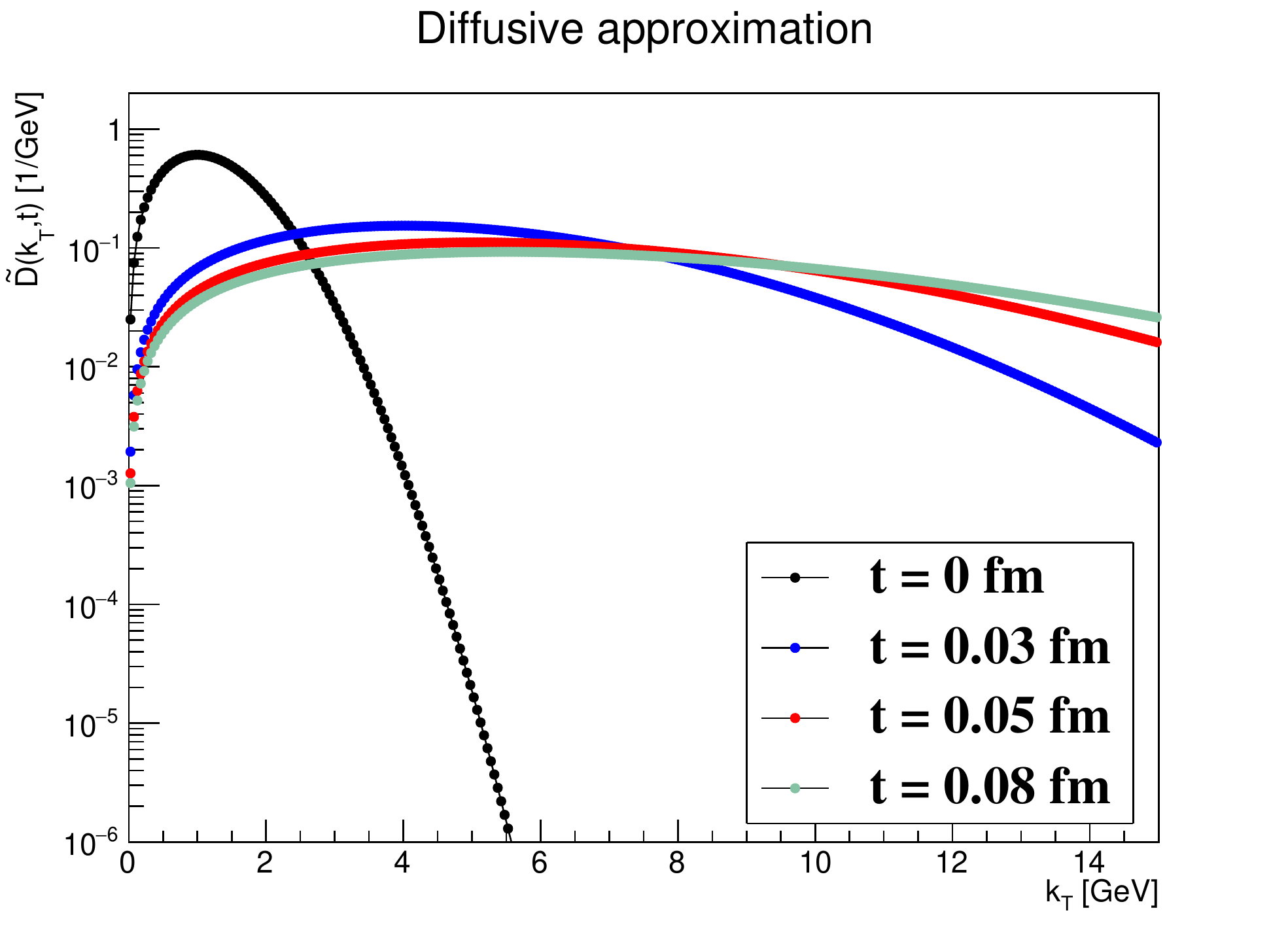}
\includegraphics[width=0.49\textwidth]{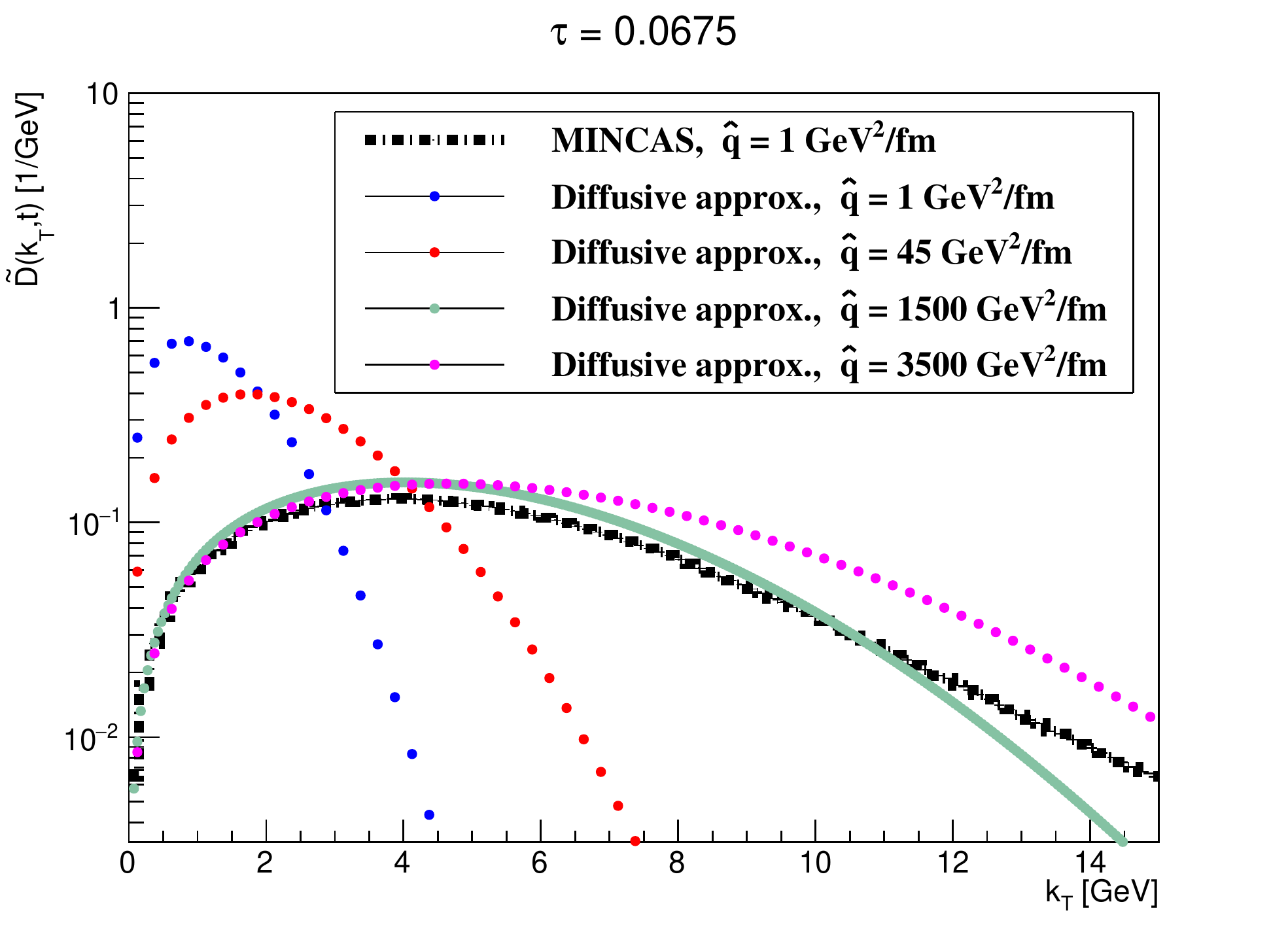}
\includegraphics[width=0.49\textwidth]{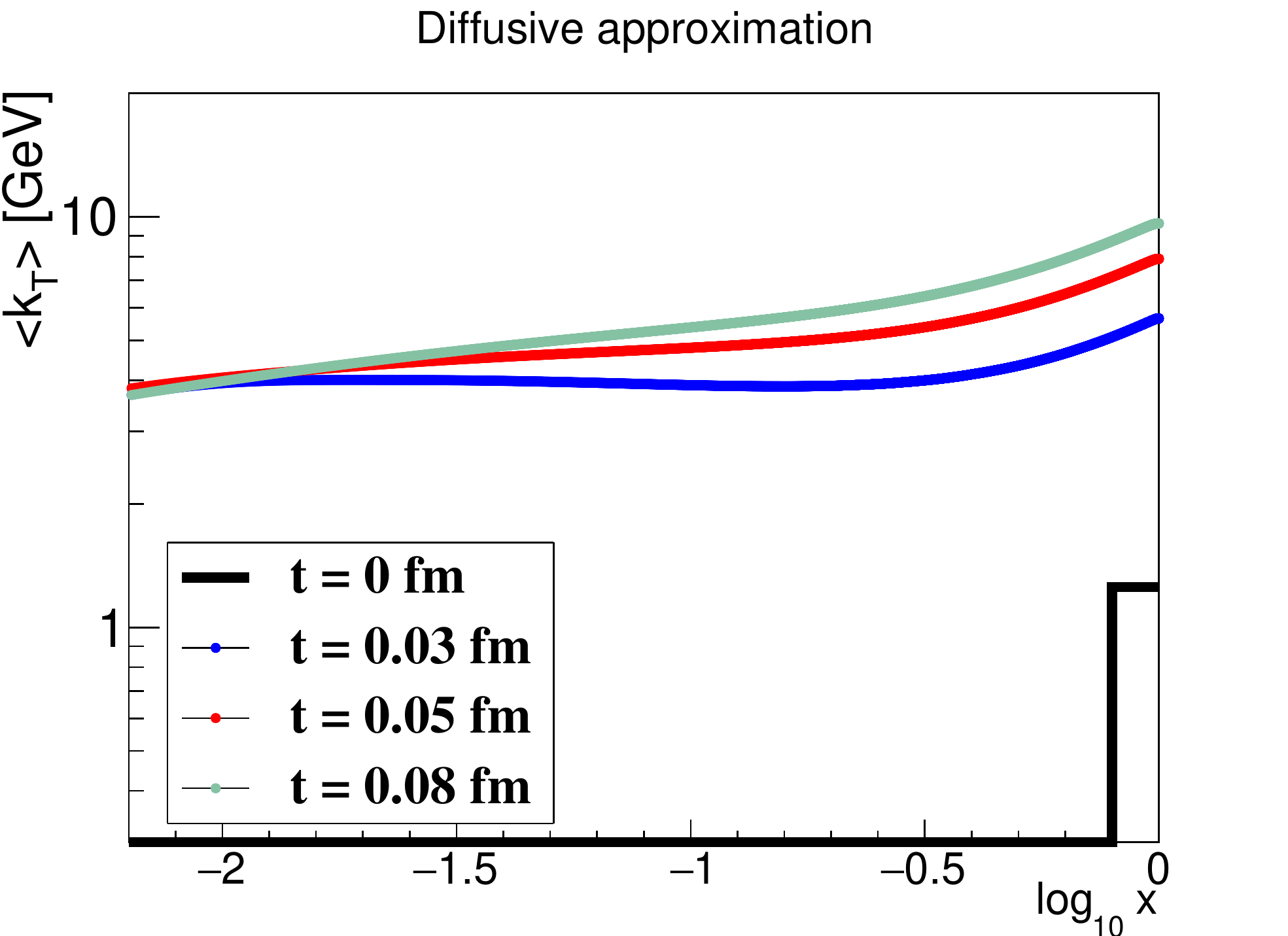}
\includegraphics[width=0.49\textwidth]{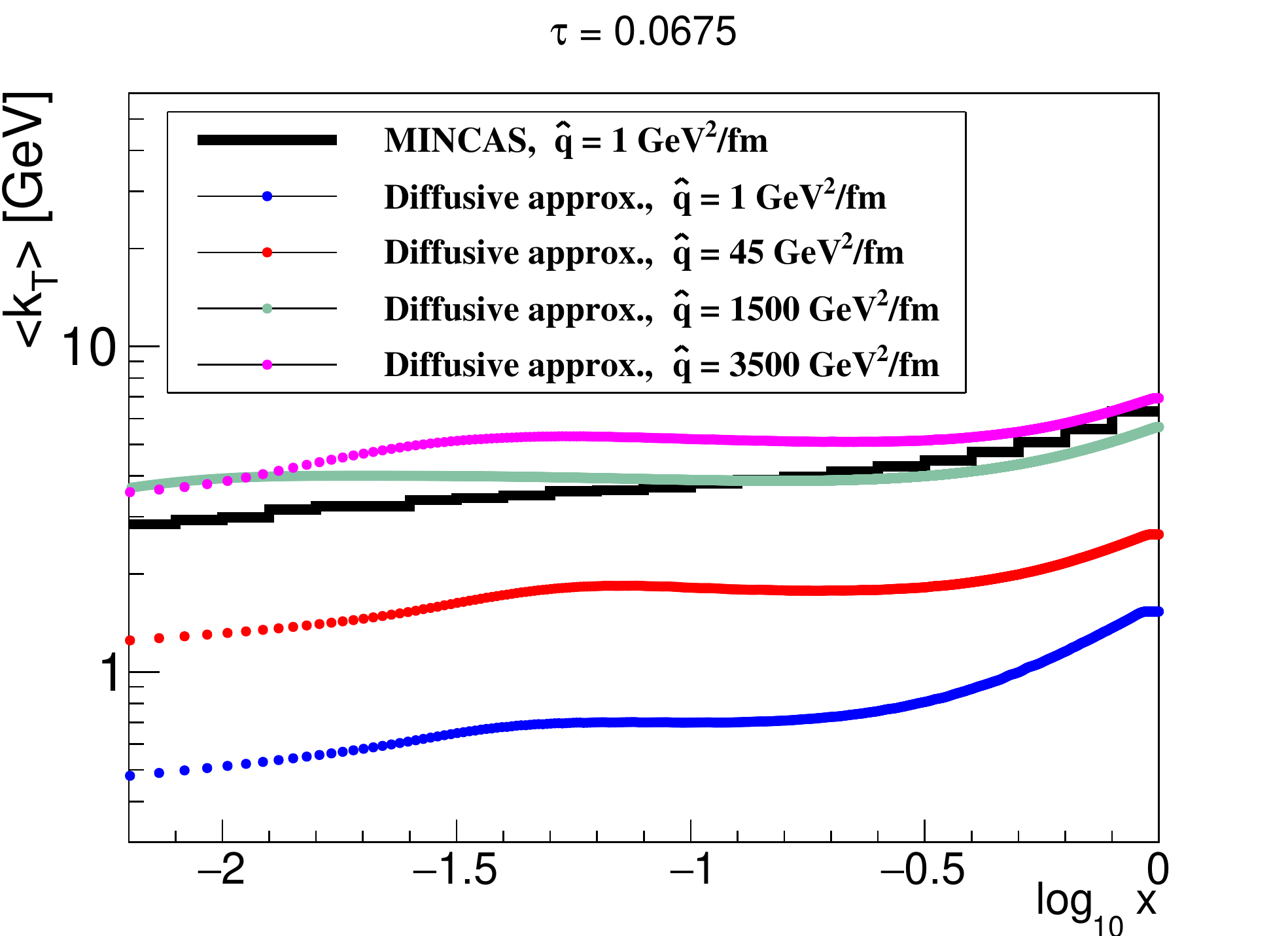}
\caption{The $k_T$ and $\langle k_T \rangle$ vs.\ $\log_{10}x$ distributions for the diffusive approximation: for four values of the evolution time $t =0,0.03,0.05,0.08\,$fm and $\hat{q}=1500\,$GeV$^2$/fm (left), and for different values of $\hat{q}$ compared with the \mincas\ results for $\hat{q}=1\,$GeV$^2$/fm and $t=1\,$fm with the $z$-only dependent kernel ${\cal K}(z)$ and the collision term of Eq.~(\ref{eq:wq1}) (right). In the diffusive approximation $\sigma_{k_0}=1\,$GeV was used; note also that the evolution times for $\hat{q}=1500\,$GeV$^2$/fm are equal to $t=0,1,2,3\,$fm ($\tau\equiv t/t^*=0.0675$ when $t=1\,$fm) in the case of $\hat{q}=1\,$GeV$^2$/fm.}
\label{fig:Robert_results}
\end{figure}

An interesting question is what is the domain of applicability of the diffusive approximation that was used in order to reduce Eq.~(\ref{eq:BDIM2}) to Eq.~(\ref{eq:ktee1_diff}). 
The approximation is advocated as a systematic expansion around $k_T$ that should be valid for rather low values of $k_T$.  However, from the explicit solution in Fig.~\ref{fig:Robert_results}  we see that the solution of the Eq.~(\ref{eq:ktee1_diff}) is reasonably reproduced in the diffusive approximation  if we allow $\hat q $ to be very large. This actually is in agreement with 
the interpretation of $\hat q$ as the average transverse momentum. Therefore, we conclude that one can describe large transverse momentum using just the diffusion approximation, but one should allow this new effective $\hat q$ to be large and different from the one in the complete equation. 
We also see that the diffusive approximation with the standard $\hat q$ preserves the general pattern of Eq.~(\ref{eq:BDIM2}),  but is much narrower than the solution of the equation before the expansion.
This feature is better visible in the plot of the $\langle k_T\rangle$ as a function of $x$ which we show for different values of $\hat q$ as well as for different values of $t$. From these results we conclude that, while the diffusive approximation is qualitatively fine, it is rather crude quantitatively.

\begin{figure}[!ht]
\centering{}
\includegraphics[width=0.32\textwidth]{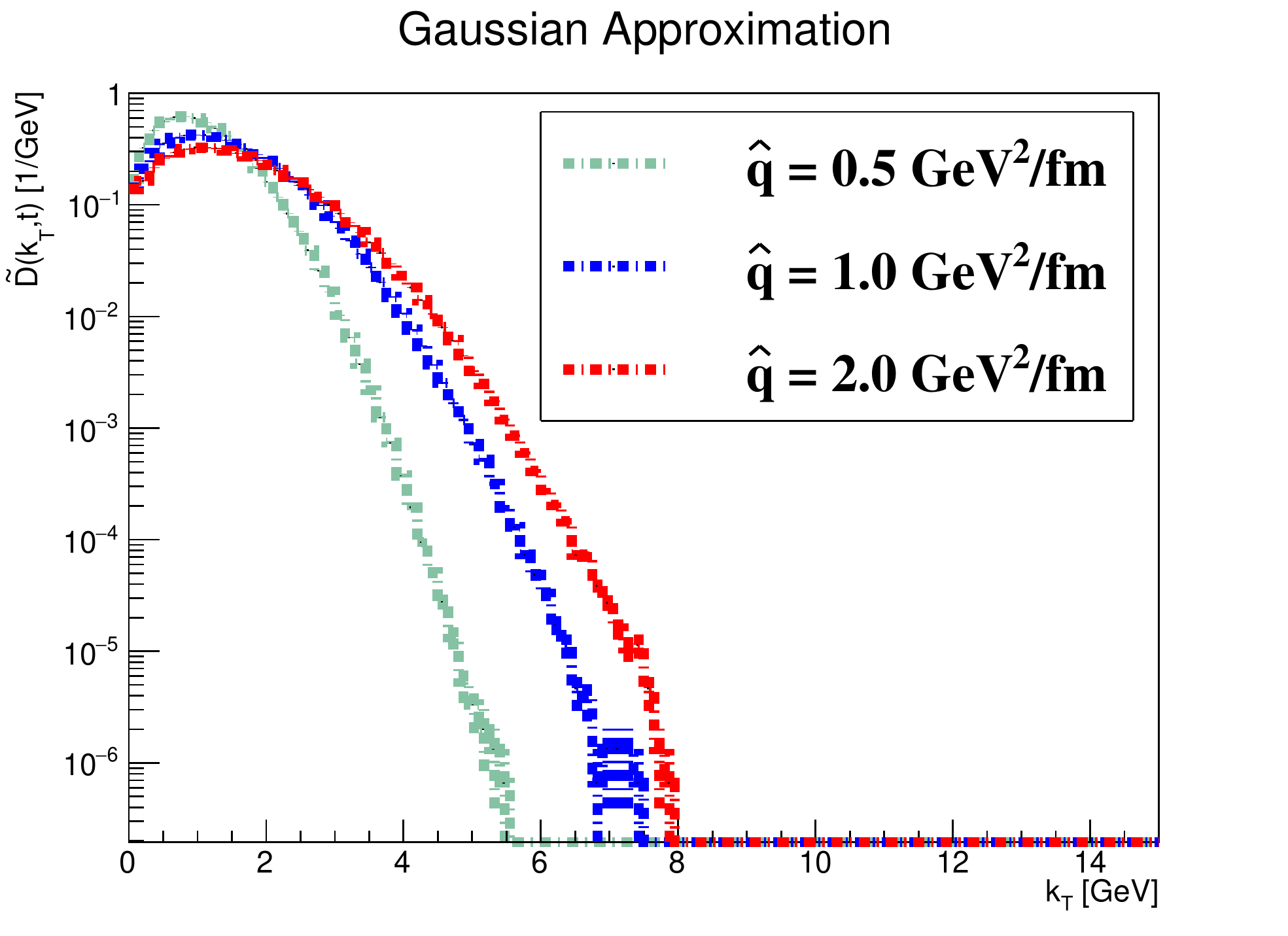}
\includegraphics[width=0.32\textwidth]{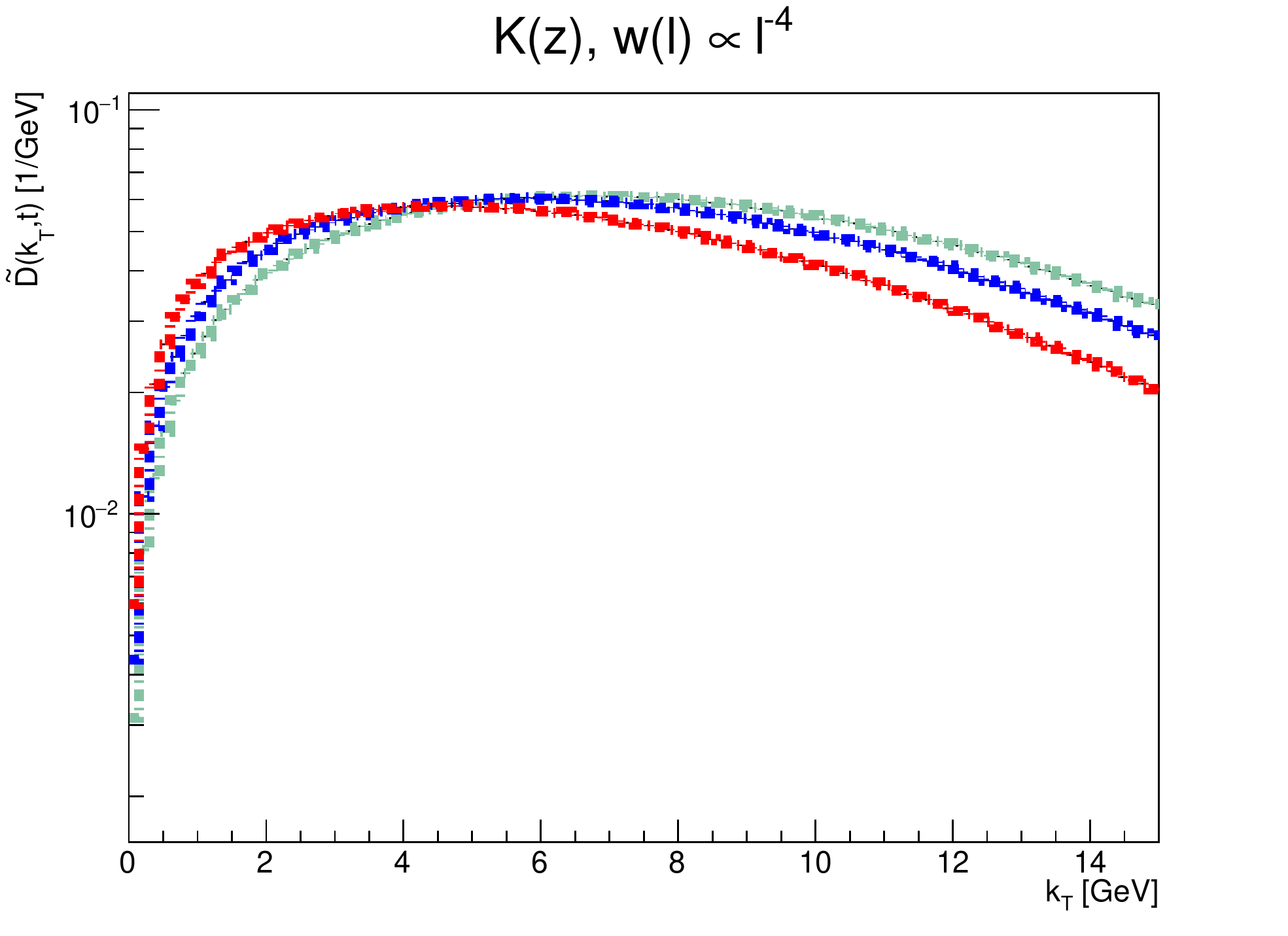}
\includegraphics[width=0.32\textwidth]{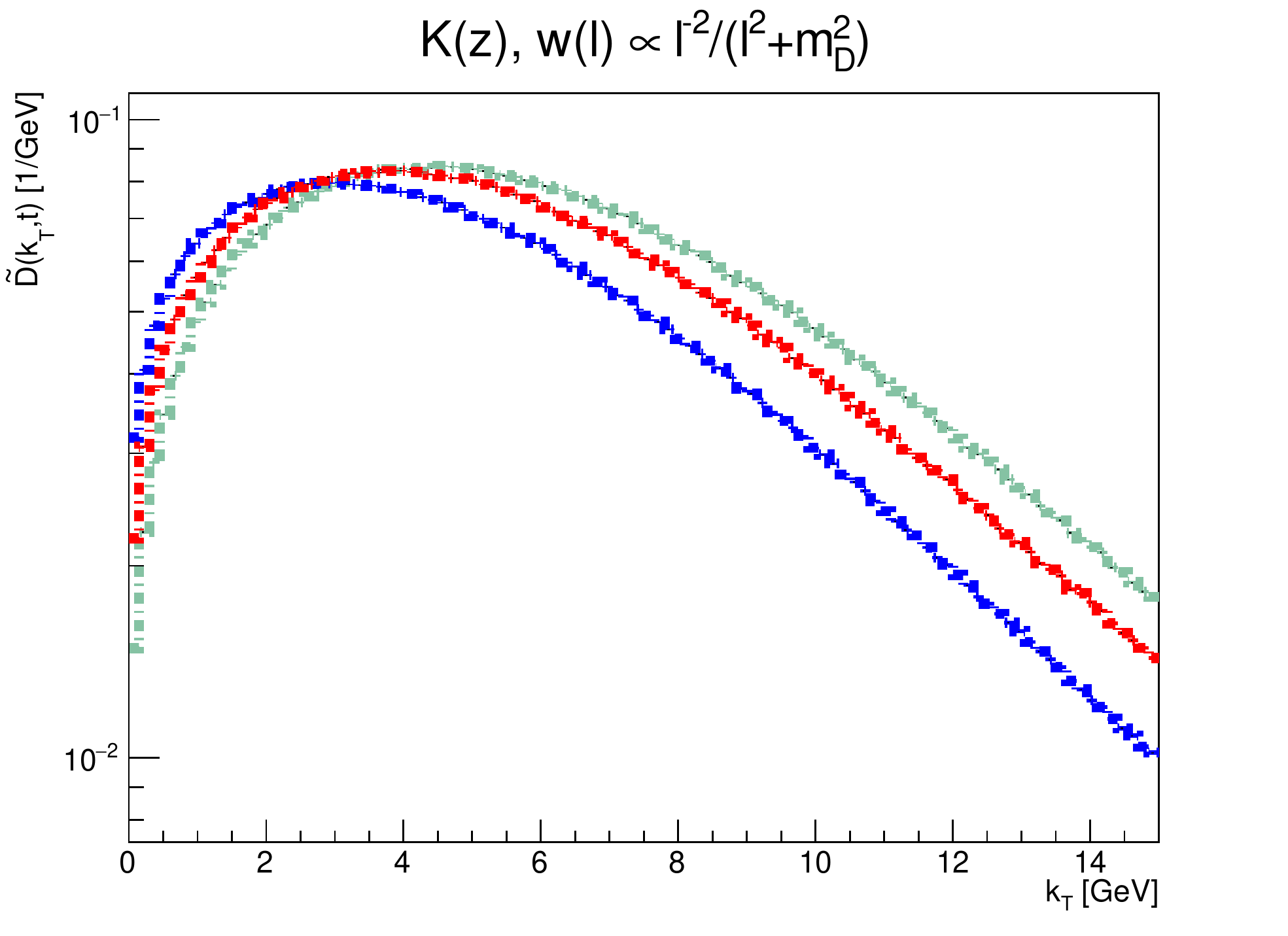}
\includegraphics[width=0.32\textwidth]{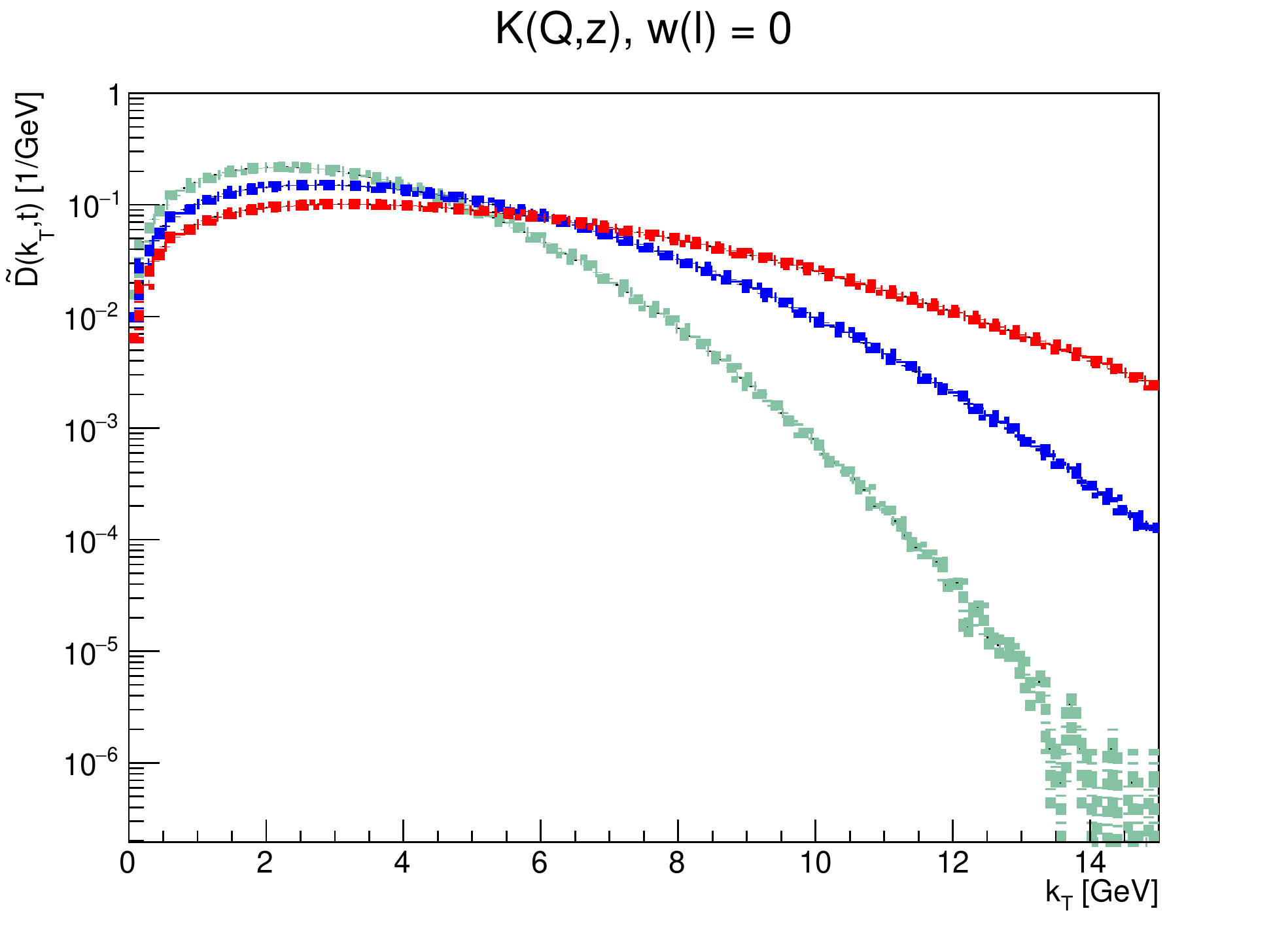}
\includegraphics[width=0.32\textwidth]{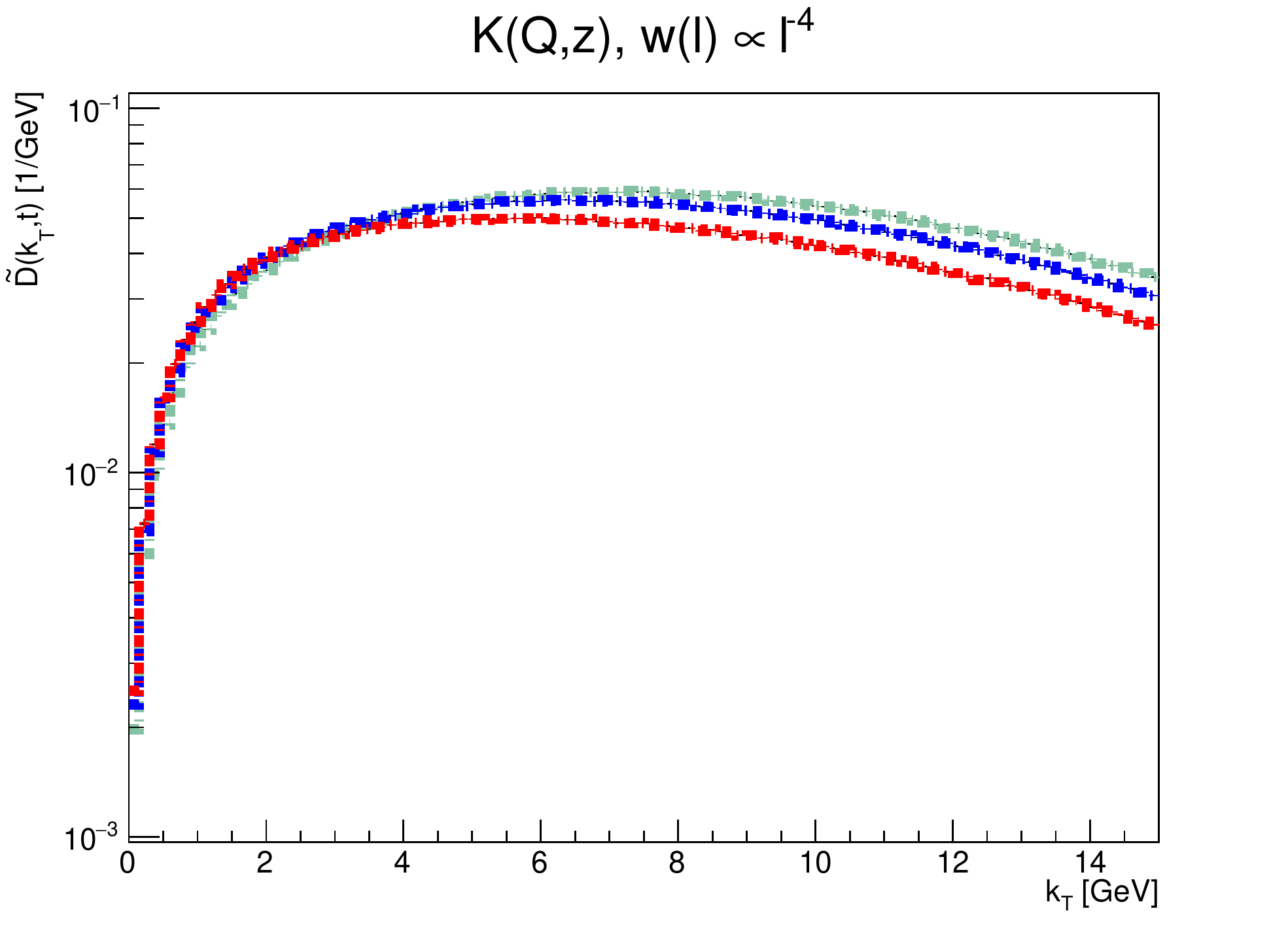}
\includegraphics[width=0.32\textwidth]{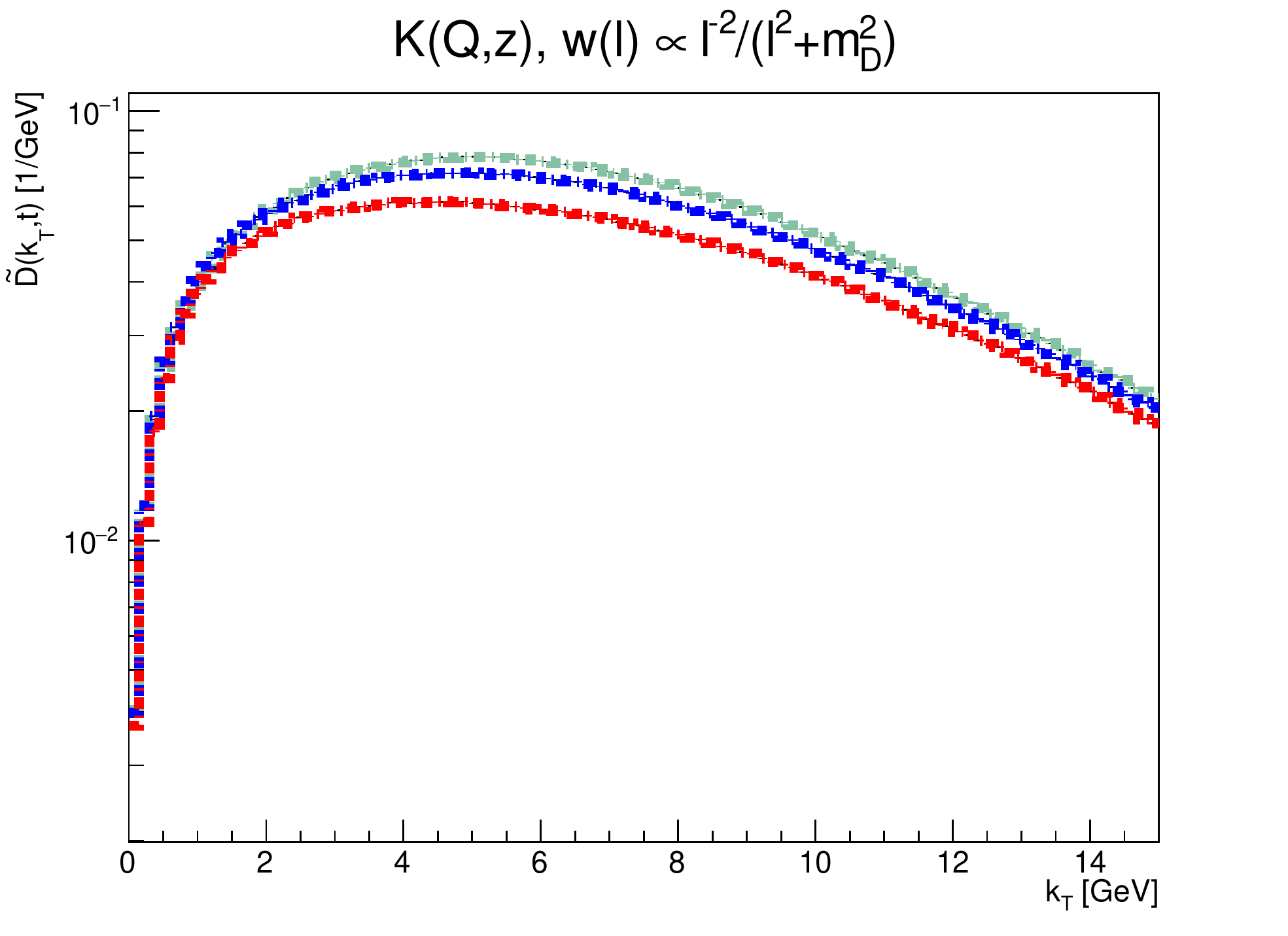}
\caption{The $k_T$ distributions for $\hat{q}=0.5,\, 1,\, 2\;$GeV$^2/$fm and $t = 4\,$fm.}
\label{fig:q_kT_distr}
\end{figure}

To complete the analysis of the $k_T$ spectrum we study on Fig. \ref{fig:q_kT_distr} its dependence on $\hat q$ for the three cases of $\hat q=0.5,\, 1,\, 2 \;$GeV$^2$/fm. We see that, in general, it is not a trivial dependence, in a sense that increasing $\hat q$ will just broaden the distribution. This is the case only for the Gaussian approximation and $w(\mathbf{l})=0$. The interpretation of this is the following. In these two cases $\hat q$ enters to some extent trivially: in the former case as a factor modifying $\langle k_{br}^2\rangle$, while in the latter in the branching term only. In the remaining cases, $\hat q$ which controls the broadening enters the branching kernel and is hidden in both the branching term and the elastic scattering term. Interplay of these two effects results in the structure visible in these cases.

\section{Conclusions and outlook}

We have solved and studied the BDIM equation as well as its various approximations, i.e. the no-momentum-transfer approximation, the diffusive approximation and the Gaussian approximation. We conclude that the momentum transfer during branching gives additional broadening that is non-negligible. Furthermore, the diffusive approximation of the elastic scattering kernel is a rather crude approximation to the BDIM equation. 
In the future it will be interesting to investigate the case of the expanding medium as well as to account for coupled evolution of quarks and gluons. Furthermore it will be interesting to see the signature of the rescattering during branching in some final state. One of the possibilities is to study decorelations of jets following \cite{vanHameren:2019xtr}.  

\section*{Acknowledgement}

This work was partially supported by the Polish National Science Centre with the grant no.\ DEC-2017/27/B/ST2/01985.
Part of the numerical computations were performed on GPUs at the Helios cluster financed by the Ministry of Education, Youth and Sports of the Czech Republic under the OP
RDE grant number CZ.02.2.67/0.0/0.0/16\_016/0002357 ``Laboratories for Excellent Bachelor and Master Degree Programmes''.

\appendix

\section{Monte-Carlo algorithm}
With the help of the Sudakov form-factor
\begin{equation}
    \Delta(p_0^+,t)=\exp{\left(-t\left[\int_{|\mathbf{q}|>q_\downarrow} \frac{d^2\mathbf{q}}{(2\pi)^2}\left( w(\mathbf{q})+\alpha_s\int_0^{1-\epsilon} dz 2z{\cal K}(\mathbf{q},z,p_0^+)\right)\right]\right)}\,,
    \label{eq:sud}
\end{equation}
where the notation $|\mathbf{q}|>q_\downarrow$ should indicate that the integration runs over all $\mathbf{q}$ except those where $|\mathbf{q}|<q_\downarrow$,
it can be shown that the following integral equation is equivalent to the integro-differential equation Eq.~(\ref{eq:BDIM1}):
\begin{align}
    D(x,\mathbf{k},t)&=D(x,\mathbf{k},t_0)\frac{\Delta(xp_0^+,t)}{\Delta(xp_0^+,t_0)}
    \nonumber\\
    &+\int_{t_0}^{t} dt'\frac{\Delta(xp_0^+,t)}{\Delta(xp_0^+,t')}\int_{|\mathbf{q}|>q_\downarrow} \frac{d^2\mathbf{q}}{(2\pi)^2}\int_0^{1-\epsilon} dz \int \frac{d^2\mathbf{Q}}{(2\pi)^2}\int_0^1 dy (2\pi)^2 
    \nonumber\\&
        \left[w(\mathbf{Q})\delta^{(2)}(\mathbf{k}-(\mathbf{Q}+\mathbf{q}) )\delta(x-y)+\alpha_s2z{\cal K}(\mathbf{Q},z,yp_0^+)\delta^{(2)}(\mathbf{k}-(\mathbf{Q}+z\mathbf{q}) )\delta(x-zy)\right]
        \nonumber\\&
        D(y,\mathbf{q},t')\,,
    \label{eq:int_eq_bdim}
\end{align}
in the simultaneous limits of $\epsilon\rightarrow 0$ and $q_\downarrow\rightarrow 0$.
 
The individual terms in Eq.~(\ref{eq:int_eq_bdim}) can be associated with probabilities:
\begin{itemize}
    \item The probability that the fragmentation function at time $t$ gets a contribution from the fragmentation function at time $t'$ without additional splitting or scattering between $t'$ and $t$ (but at $t'$ and $t$ some particle interaction occurs):
    \begin{equation}
        \frac{\Delta(xp_0^+,t)}{\Delta(xp_0^+,t')}.
    \end{equation}
    \item The probability density that the fragmentation function at the momentum fraction $x$ and the transverse momentum $\mathbf{k}$ gets a contribution from the fragmentation function at the earlier time $t'$ at the momentum fraction $y$ and the transverse momentum $\mathbf{q}$ via a splitting with momentum fraction $z$ and transverse momentum $\mathbf{Q}$, where $x=zy$ and $\mathbf{k}=\mathbf{Q}+z\mathbf{q}$:
    \begin{equation}
        \frac{z{\cal K}(\mathbf{Q},z,yp_0^+)}{\int d^2{\mathbf{Q}}\int_0^{1-\epsilon} dzz{\cal K}(\mathbf{Q},z,yp_0^+)}\,.
        \label{eq:selsplit1}
    \end{equation}
    Thus, the probability for a splitting with a certain $z$ value (independent of the value of $Q$) is given as
        \begin{equation}
        \frac{z{\cal K}(z)}{\int_0^{1-\epsilon} dz'z'{\cal K}(z')},
        \label{eq:zdist}
    \end{equation}
    where  ${\cal K}(z)$ is 
    \begin{equation}
       {\cal K}(z)=\int d^2\mathbf{Q}  {\cal K}(\mathbf{Q},z,yp_0^+)\frac{\sqrt{yp_0^+}}{2\pi\sqrt{\hat{q}}}=\frac{f(z)^{5/2}}{(z(1-z))^{3/2}}\,.
       \label{eq:fromKzQtoKz}
    \end{equation}
    \item The probability density that the fragmentation function at the transverse momentum $\mathbf{k}$ gets a contribution from the fragmentation function at the earlier time $t'$ at  the transverse momentum $\mathbf{q}$ via a scattering with the transverse momentum $\mathbf{Q}$, where  $\mathbf{k}=\mathbf{Q}+\mathbf{q}$:
    \begin{equation}
        \frac{w(\mathbf{Q})}{\int_{|\mathbf{Q'}|>q_\downarrow} d^2\mathbf{Q}' w(\mathbf{Q'})}\,.
    \end{equation}
    %
\end{itemize}
Thus, it is possible to obtain solutions for Eq.~(\ref{eq:BDIM1}) via a Monte-Carlo algorithm, 
where a distribution $D(x,\,\mathbf{k},\,t)$ that obeys Eq.~(\ref{eq:int_eq_bdim}) can be obtained by selecting independently of one another a large number $N_{ev}$ of sets $(x,\,\mathbf{k})$, which follow $D(x,\,\mathbf{k},\,t)$.

In each of the $N_{\rm ev}$ cases, the $x$ and $\mathbf{k}$ values are obtained in the following way: 
\begin{itemize}
    \item Some initial values $x_0$, $\mathbf{k}_0$ are set together with the time $t_0$ of the start of the evolution.
    \item For every set $(x_i,\,\mathbf{k}_i,\,t_i)$, $i\in \mathbb{N}$, a new set $(x_{i+1},\,\mathbf{k}_{i+1},\,t_{i+1})$ is selected, where $t_{i+1}>t_{i}$.
    \item The previous step is repeated until for some time $t_j$ $j\in \mathbb{N}$, it is found that $t_j\geq t$. 
    Then the algorithm gives $x=x_{j-1}$, $\mathbf{k}_j=\mathbf{k}_{j-1}$ and stops.
\end{itemize}

The selection of a set $(x_{i+1},\,\mathbf{k}_{i+1},\,t_{i+1})$ from a set $(x_i,\,\mathbf{k}_i,\,t_i)$ is done in the following way:
\begin{enumerate}
        \item Select time $t_{i+1}$ of next splitting/scattering by first choosing a random number $R\in[0,1]$ from a uniform distribution and then solving the equation 
        \begin{equation}
          R=\frac{\Delta(xp_0^+,t_{i+1})}{\Delta(xp_0^+,t_{i})}\,.  
        \end{equation}
        The result of this calculation is
        \begin{equation}
            t_{i+1}=t^\ast\left(\frac{t_i}{t^\ast}-\frac{\ln(R)}{\int_0^{1-\epsilon} dzz{\cal K}(z)\frac{1}{\sqrt{x_{i}}}+t^\ast \int_{|\mathbf{q}|>q_\downarrow}\frac{d^2\mathbf{q}}{(2\pi)^2}w(\mathbf{q})}\right)\,.
            \label{eq:selt}
        \end{equation}
        \item Determine whether a splitting or scattering occurs:\\
        This is done, by first selecting a random number $R\in[0,1]$ from a uniform distribution.
        If 
        \begin{equation}
            R>\frac{
            \int_0^{1-\epsilon} dzz{\cal K}(z)\frac{1}{\sqrt{x_{i}}}}{\int_0^{1-\epsilon} dzz{\cal K}(z)\frac{1}{\sqrt{x_{i}}}+t^\ast \int_{|\mathbf{q}|>q_\downarrow}\frac{d^2\mathbf{q}}{(2\pi)^2}w(\mathbf{q})}
        \end{equation}
        a scattering occurs, otherwise a splitting.
        \item If a splitting occurs, determine $x_{i+1}$ and $\mathbf{k}_{i+1}$ as follows:
        \begin{enumerate}
            \item Select $z$ from $\mathcal{K}(z)$ by choosing a random number $R\in[0,1]$ from a uniform distribution and then solve the equation 
            \begin{equation}
                R=\frac{\int_0^z dz'z'{\cal K}(z')}{\int_0^{1-\epsilon} dz''z''{\cal K}(z'')}.
                \label{eq:selz}
            \end{equation}
            This equation is solved approximately by first tabulating values of $\int_0^z dz'z'{\cal K}(z')$ for a set of $z$ values that is sufficiently dense for the desired accuracy and then searching from this table the $z$ value, which is the closest to the one that solves Eq.~(\ref{eq:selz}).
            \item Select $Q$ from $\mathcal{K}(Q,z)$ by choosing random number $R\in[0,1]$ from a uniform distribution and solving for $a:=\frac{Q^2}{2k_{br}^2}$ the equation
                \begin{equation}
                    R=\frac{\int_0^a da' \sin(a')e^{-a'}}{\int_0^\pi da'' \sin(a'')e^{-a''}}=\frac{1-(\cos(a)+\sin(a))e^{-a}}{1+e^{-\pi}}\,.
                \label{eq:selQsplit}
                \end{equation}
            After selection of $a$, the value of $Q=\sqrt{2k_{br}^2a}$ is calculated.
            While the values of $a$ can assume any positive value, we here constrain the values to the region $a\in[0,\pi]$ in order to avoid the region where $\sin(a)e^{-a}$ becomes negative.
            Indeed the splitting function in the form of Eq.~(\ref{eq:Kqz}) was deduced in Ref.~\cite{Blaizot:2013vha} in the harmonic approximation, which needs corrections at large momentum scales.
            \item Select the azimuthal angle $\phi\in [0,\,2\pi ]$ from a uniform distribution.
            \item Obtain $x_{i+1}$ as $x_{i+1}=x_iz$.
            \item Obtain $\mathbf{k}_{i+1}$ as $\mathbf{k}_{i+1}= \mathbf{Q}+z \mathbf{k}_i$ via
            \begin{align}
                k_{i+1,x}&=Q\,\cos\phi+zk_{i,x}\,,\\
                k_{i+1,y}&=Q\,\sin\phi+zk_{i,y}\,,
            \end{align}
            where the subscripts $x$ and $y$ denote the respective Cartesian coordinates of the momenta $\mathbf{k}_i$ and $\mathbf{k}_{i+1}$.
        \end{enumerate}
    \item If a scattering occurs, determine $\mathbf{k}_{i+1}$ as follows:
        \begin{enumerate}
            \item Select $Q$ by choosing from a uniform distribution a random value $R\in[0,1]$ and then solving for $Q$ the equation
            \begin{equation}
                R=\frac{\int_{q_\downarrow}^Q d^2Q'w(\mathbf{Q'})}{\int_{q_\downarrow}^\infty d^2Q'' w(\mathbf{Q''})}\,.
            \end{equation}
            For the scattering kernel of the form given in Eq.~(\ref{eq:wq1}), this equation has the following solution:
            \begin{equation}
                Q=\frac{q_\downarrow}{\sqrt{1-R}}\,.
                \label{eq:selQscat}
            \end{equation}
            \item Obtain $\mathbf{k}_{i+1}$ as $\mathbf{k}_{i+1}=\mathbf{Q}+\mathbf{k}_{i}$.
        \end{enumerate}

\end{enumerate}
%

\section{Deterministic method}

Eq.~(\ref{eq:ktee1_diff}) can be rewritten in the polar coordinates as:
\begin{equation}
\begin{aligned}
\frac{\partial}{\partial t} D(x,k,\phi,t) = & \: \frac{1}{t^*} \int_0^1 dz\, {\cal K}(z) \left[\frac{1}{z^2}\sqrt{\frac{z}{x}}\, D\left(\frac{x}{z},\frac{k}{z},\phi,t\right)\theta(z-x) 
- \frac{z}{\sqrt{x}}\, D(x,k,\phi,t) \right] \\
+& \hat q\frac{1}{4}\left[\left(\frac{\partial}{\partial k}\right)^2+\frac{1}{k}\frac{\partial}{\partial k}+\frac{1}{k^2}\frac{\partial}{\partial\phi }\right] D(x,k,\phi,t).
\end{aligned}
\label{eq:ktee2}
\end{equation}
The initial condition for the $D(x,k,\phi,t)$ is given by
\begin{equation}
D(x,k,\phi,0) =
\begin{cases}
     \frac{1}{2\pi\sigma^2}\exp{\left(-\frac{k^2}{2\sigma^2}\right)} & \quad \text{for } x=1,\\
     0 & \quad \text{for } 0\leq x<1,
\end{cases}
\label{eq:initcond}
\end{equation}
where $\sigma=1\,$GeV. The equation is symmetric with respect to the polar angle $\phi$, so the corresponding Laplacian simplifies.

In order to get the integrated distribution one needs to calculate the integral:
\begin{equation}
D(t,x)=\int d\phi\,dk\,k\,D(t,x,k,\phi)   
\end{equation}
The equation can be solved directly for the $\phi$-integrated distribution, since the $\phi$-dependence is trivial.

The terms on RHS of the Eq.~(\ref{eq:ktee2}) are evaluated by central differences (the Laplacian of $k$ with one-sided approximations at the boundaries of the computational domain) and by the box-rule (the integral term):
\begin{equation}
\begin{aligned}
    \frac{ \partial D_{i,j}(t)}{\partial t} = & \: \frac{\hat{q}}{4}\left( \frac{1}{2k_j\Delta k}\left(D_{i,j+1}(t)-D_{i,j-1}(t)\right) + \frac{1}{(\Delta k)^2}\left(D_{i,j+1}(t) - 2D_{i,j}(t) + D_{i,j-1}(t)\right)\right)\\
    +& \frac{1}{t^*} \sum_{l=i}^{N_x} \Delta x\, {\cal K}(x_l) \left[\frac{1}{x_l^2}\sqrt{\frac{x_l}{x_i}}\, D_{(i/l,j/l)}(t) 
- \frac{x_l}{\sqrt{x_i}}\, D_{i,j}(t) \right].
\end{aligned}
\label{eq:rkrhs}
\end{equation}

A numerical grid is equidistant and 2-dimensional (we drop the $\phi$-dependence due to the symmetry of the problem):
\begin{equation}
    x_i = i\Delta x,\quad k_j= j\Delta k,\quad i\in[0,N_x-1],\quad j\in[0,N_k-1],\quad \Delta x=\frac{1}{N_x},\quad \Delta k=\frac{k_{max}}{N_k}.
\end{equation}

We solve Eq.~(\ref{eq:rkrhs}) to obtain the functions $D_{i,j}(t_n)=D(x_i,k_j,t_n)$ at given points $x_i$, $k_j$ and a time level $t_n$. The initial condition is given by Eq.~(\ref{eq:initcond}). The number of grid points for $x$ and $k$ is increased up to $N_x=10240$ and $N_k=1000$ with $x\in[0,1]$ and $k\in[0,50]$ ($k_{max}=50$) for the case of $\hat{q}=1500\,$GeV$^2$/fm, for other $\hat{q}$ we used coarse grid with $N_x=1024$ and $N_k=200$.

We use a fourth-order Runge--Kutta method to obtain the numerical solution of the Eq.~(\ref{eq:rkrhs}) in time (the Cash--Karp method with the adaptive time stepping \cite{Cash:1990} is employed). The time step is being changed according to the following formula:
\begin{equation}
\Delta t =
\begin{cases}
     0.9\Delta t\left(\frac{\rm TOL}{\rm E}\right)^{0.2} & \quad \text{for } {\rm E} <{\rm TOL},\\
     0.9\Delta t\left(\frac{\rm TOL}{\rm E}\right)^{0.25} & \quad \text{for } {\rm E}\geq {\rm TOL},
\end{cases}
\end{equation}
where ${\rm TOL}=10^{-6}$ is a tolerance and $\rm E$ is the maximal error in the last step of the embedded Runge--Kutta method.
In order to minimise the computational time, the numerical code was parallelized and implemented in NVIDIA CUDA (double precision was used in computations).

\bibliography{refs}{}
\bibliographystyle{jhep}

\end{document}